\documentclass[book]{agujournal2019}
\usepackage{lineno}
\usepackage[inline]{trackchanges} 
\usepackage{soul}

\usepackage[utf8]{inputenc}

\usepackage{parskip}
\usepackage{graphicx}
\usepackage[inline]{enumitem}
\usepackage{natbib}
\usepackage{amsmath,bm}
\usepackage{xfrac}
\usepackage[bb=boondox]{mathalfa}
\usepackage[inline]{enumitem}
\usepackage[justification=centering]{caption}
\usepackage[export]{adjustbox}
\usepackage{float}
\usepackage{natbib}
\usepackage{array, makecell}
\usepackage{multirow}
\usepackage{algorithm2e}
\usepackage{hyperref}
\hypersetup{
    colorlinks=true,
    linkcolor=blue,
    filecolor=magenta,      
    urlcolor=cyan,
    citecolor=blue
    }
\setcitestyle{square}
\graphicspath{{Images/}}
\DeclareMathOperator{\arctantwo}{arctan2}

\journalname{JGR: Machine Learning and Computation}

\begin{document}

%
%

\title{Modeling groundwater levels in California's Central Valley by hierarchical Gaussian process and neural network regression}

%
%




\authors{Anshuman Pradhan\affil{1}, Kyra H. Adams\affil{2}, Venkat Chandrasekaran\affil{1}, Zhen Liu\affil{2}, John T. Reager\affil{2}, Andrew M. Stuart\affil{1} and Michael J. Turmon\affil{2}}


\affiliation{1}{Computing and Mathematical Sciences, California Institute of Technology}
\affiliation{2}{Jet Propulsion laboratory, California Institute of Technology}




\correspondingauthor{Anshuman Pradhan}{pradhan1@alumni.stanford.edu}




\begin{keypoints}
\item Groundwater levels in California's Central Valley are sampled irregularly and noisily making groundwater management challenging.
\item Machine learning methodology of hierarchical Gaussian process and neural network regression is formulated to model groundwater levels.
\item Modeled uncertainty estimates of 2015-2020 groundwater levels are validated to be consistent with the data distribution of 90 blind wells.
\end{keypoints}
%
%

%
%


\begin{abstract}
Modeling groundwater levels continuously across California's Central Valley (CV) hydrological system is challenging due to low-quality well data which is sparsely and noisily sampled across time and space. The lack of consistent well data makes it difficult to evaluate the impact of 2017 and 2019 wet years on CV groundwater following a severe drought during 2012-2015. A novel machine learning method is formulated for modeling groundwater levels by learning from a 3D lithological texture model of the CV aquifer. The proposed formulation performs multivariate regression by combining Gaussian processes (GP) and deep neural networks (DNN). The hierarchical modeling approach constitutes training the DNN to learn a lithologically informed latent space where non-parametric regression with GP is performed. We demonstrate the efficacy of GP-DNN regression for modeling non-stationary features in the well data with fast and reliable uncertainty quantification, as validated to be statistically consistent with the empirical data distribution from 90 blind wells across CV. We show how the model predictions may be used to supplement hydrological understanding of aquifer responses in basins with irregular well data. Our results indicate that on average the 2017 and 2019 wet years in California were largely ineffective in replenishing the groundwater loss caused during previous drought years.
\end{abstract}
\section*{Plain Language Summary}
Building a reliable model of groundwater level depths in California's Central Valley (CV) aquifer system is essential for groundwater management and decision-making. However, publicly available water level well data are sparse, irregular, and noisy, resulting in large uncertainties in groundwater modeling efforts. We mathematically formulate a novel machine learning approach, termed the Gaussian process and deep neural network (GP-DNN) regression, to constrain the uncertainties on groundwater levels in CV with input information from a model of the aquifer lithology. The machine learning model uses DNNs to extract useful features from the aquifer lithological model. Subsequently, GP regression approach conducts interpolation in the lithological feature space to predict water levels at every location in the CV, as well as provide rigorous estimates of modeling and data uncertainty. The model uncertainty predictions were validated to be reliable by statistically comparing results at 90 wells in the study area that were kept blind during the modeling process. We show how proposed machine learning method may be used to overcome data limitation challenges in hydrological basins and improve understanding of aquifer response to groundwater recharge and drought recovery.

%
%

%


%
%
%
%

\section{Introduction} \label{introduction}
With climate change causing frequent periods of intense droughts in many parts of the world, the need for sustainably managing groundwater resources has never been greater. A prime example is California's Central Valley (CV) aquifer system. The CV, supporting a 20 billion dollar per year agricultural industry (\cite{faunt16}), has been severely strained by recent droughts, depleting groundwater levels and diminishing surface water availability. One of the key variables impacting decisions in groundwater management is understanding how water levels in the underlying aquifer system vary through processes of groundwater recharge and discharge, such as precipitation and agricultural pumping. A key challenge in the CV is the absence of water level measurements that are regularly sampled across space and time. Publicly available groundwater level databases often do not include data from privately-owned wells in the valley (\cite{johnson15}, \cite{kim21}). Available measurements are often noisy and sampled with large temporal gaps (see section \ref{Application} for additional discussion). These factors impede developing a comprehensive understanding of how CV hydrogeology and groundwater processes such as recharge, depletion, and aquifer subsidence are interlinked, rendering decision-making for water resources management difficult.

A traditional approach to comprehensively understanding subsurface water levels is to model the groundwater level or hydraulic head of the aquifer as the response of a physics-based groundwater flow model (\cite{todd05}, \cite{harbaugh05}, \cite{faunt09}). 3D simulation of groundwater flow requires a conceptual model of the aquifer and quantification of rock hydraulic and storage properties. Since subsurface properties are not observed exhaustively, it becomes necessary to derive them inversely to be able to perform flow simulation. In addition to head measurements at wells, remote sensing (\cite{chaussard14},  \cite{liu19}) and hydrogeophysical data may be used in the inversion process (\cite{binley15}, \cite{smith19}, \cite{kang21}). Given the sparse and noisy nature of available water data in CV, it is highly desirable to quantify any uncertainty associated with the modeling process. If the prior estimates of uncertainty are modeled by probability distributions, the Bayesian inversion paradigm may then be employed to condition the prior uncertainty to observed data (\cite{tarantola05}, \cite{stuart10}). A large source of uncertainty for subsurface property inversion relates to the model parameterization (\cite{caers11}), for instance specifying the structure of faults and the stratigraphy of aquifer confining units, as well as the spatial distribution of lithofacies and intra-facies aquifer property variability.

Over the past few decades, research developments in the geostatistics literature have led to the development of several sophisticated models for spatial heterogeneity associated with subsurface reservoirs, including covariance based, training image based, object-based, surface-based and process-mimicking models (\cite{deutsch98}, \cite{caers05}, \cite{mariethoz14}, \cite{pyrcz14}). Covariance-based models include variants of kriging or Gaussian process regression method, which enforce constraints on the two-point correlations or second order statistics in the modeling domain. Constraining just the spatial covariance model has been often found to be limited in replicating the complex geological heterogeneity associated with common geological settings, for instance a fluvial aquifer system containing channelized sand lithofacies (\cite{feyen06}, \cite{linde2015}). To address these challenges, several methods for modeling higher-order spatial correlations have been developed, a few of which were listed previously. However, such models often rely on techniques from computer-vision to stitch together complex spatial patterns or drop geological objects into a modeling grid, and are not amenable for conditioning to dense geophysical or remote-sensing data (\cite{bertoncello11}, \cite{pradhan20a}), unlike kriging-based approaches. Stochastic search methods, such as Markov Chain Monte-Carlo (MCMC), are often necessitated for model inversion with uncertainty quantification (\cite{mariethoz10}, \cite{keating10}, \cite{laloy13}, \cite{hermans15}). Stochastic search methods are known for their computational complexity, especially when optimizing for 3D aquifer models with millions of grid cells.

Given the several limitations associated with subsurface modeling, we adopt a data-driven approach to groundwater modeling and uncertainty quantification. In many cases, the final groundwater management decisions are not directly dependent on the earth model itself, but rather on some summary statistics or trends of groundwater variability. For instance, water level long-term and seasonal variability trends may be used to understand aquifer response to groundwater recharge and discharge (\cite{riel2018}, \cite{neely2021}), aiding groundwater budgeting efforts during dry years. In this paper, we seek to estimate these trends continuously across CV using sparse well data. We propose a novel methodology combining Gaussian process (GP) regression and deep neural networks (DNN) to achieve this objective. Gaussian process regression, also known as kriging, was introduced almost seventy years ago by \cite{krige51}, with subsequent theoretical development made by \cite{matheron62}. Since then, kriging and its several variants have seen widespread usage in geosciences and spatial statistics (\cite{journel78}, \cite{cressie93}, \cite{goovaerts97}). Our primary motivation for employing the GP regression methodology is that it allows fast and easy derivation of uncertainty estimates. However, a fundamental limitation of kriging is its assumption of spatial stationarity across the modeling domain. This limits the ability to model non-stationary data, for instance data with varying spatial length-scales across different regions.

In geostatistics literature, non-stationarity has been handled with techniques such as kriging with locally varying mean, universal kriging and intrinsic random functions (\cite{marsily87}). Kriging has received emerging interest in the machine learning literature where it is more commonly known as GP regression (\cite{rasmussen06}).  This has led to the development of several sophisticated mathematical formulations for handling non-stationary data with GPs. Two broad categories of developments may be identfied. In the first category, the non-stationarity challenge is addressed by construction of hierarchical formulations of covariance kernels. \citet{paciorek03} propose to model non-stationary data by GPs whose covariance function depends on another GP. \citet{damianou13} proposed the methodology known as deep Gaussian processes, where each GP layer with stationary covariance, is composed from another stationary GP. \cite{dunlop18} extend \citet{paciorek03}'s work on multiple hierarchical layers of covariance functions and propose to handle non-stationarity by iteratively modifying the length-scales of covariance functions. In \citet{roininen19}'s paper, the hierarchy is built using the stochastic partial differential equation representation of GP (\cite{lindgren11}). While these complex kernels have some shown promising results, they lack the representational capabilities in high dimensions that has become commonplace in machine learning with DNNs (\cite{bradshaw2017}). The other category involves applying standard GP regression in a latent space obtained by deformation of the input feature space such that the assumptions of stationarity hold in this latent space. \citet{sampson92} proposed obtaining the latent space by warping the geographical coordinate space. Specifically, multidimensional scaling (MDS) of the training data was conducted such that empirical estimates of the variogram were preserved during MDS, and subsequently spline mappings were used to derive a smooth latent field in the MDS space. In contrast to two step modeling of the latent space, \citet{schmidt03} proposed a fully Bayesian approach that involves performing the spatial deformation by Gaussian process priors. However, their approach requires expensive MCMC algorithms to sample the posterior. \cite{sampson01} provide a review of some of the original works involving warping of the input space. Recently, there has been a resurgence of interest in this approach, especially on modeling the latent space by DNNs. \cite{calandra2016} and \cite{wilson2016}  perform GP regression in the latent space learned with neural networks, referring to their approaches as manifold GP regression and deep kernel learning respectively.  \cite{bradshaw2017} apply the GP hybrid DNN model to image classification task. It was demonstrated how deep neural networks boosted the capability of GP to model non-stationary, discontinuous and noisy data. Note that the above approaches were formulated in the univariate regression or classification setting. 

In this paper, we extend the above formulation to regression in the multivariate setting with geospatially and hydrogeologically indexed data. By hydrogeological indexing, we refer to the lithological data that was also used as input features to the model in addition to geospatial coordinates. We establish a two-level model hierarchy with a DNN below a GP layer, trained end-to-end. As shown in section \ref{Application}, proposed model is able to handle the non-stationary, sparse and noisy well data by learning to appropriately scale the coordinates of the latent space. The model allows analytical derivation of the posterior uncertainty and fast generation of posterior samples. The intended novel contributions of this paper are as follows.
\begin{enumerate}
    \item We extend the formulation of manifold GP regression or deep kernel learning to handle geospatially and hydrogeologically indexed data with multivariate prediction variables. The novel methodology is simply referred to as GP-DNN regression.
    \item We formulate a novel cross-validation technique using chi-square quantile-quantile plots for evaluating generalization capability of GP-based regression models. 
    \item We present a real world case study of groundwater level modeling in CV by GP-DNN regression. Specifically, we assume a linear model for seasonal and long-term variability of groundwater levels and demonstrate how the machine learning methodology may be used with this groundwater model, and irregular datasets to yield statistically valid uncertainty estimates without detailed physics-based modeling. 
\end{enumerate}

 A theoretical description of the proposed methodology, followed by the real world case study from California's CV is presented next. The application is focused around modeling long-term and seasonal trends in groundwater levels in  CV from 2015-2020. We also provide interpretations and visualizations of the latent space learned by the DNN, explicitly demonstrating how it handles non-stationarity and uncertainty. We present hydrological interpretations about the seasonal and long-term trends of water levels in CV, especially in the context of drought and recovery to illustrate how this method may be applied to real-life understanding of hydrologic data. Finally, we discuss future work, both in terms of how the methodology maybe extended to handle more complicated real-world data noise scenarios as well how the CV groundwater model may be made more rigorous by employing the methods presented in this paper.

\section{Methodology} \label{methodology}
Let $\mathcal{D} \in \mathbb{R}^2$ denote the 2D geospatial domain of the area of interest. Let $\bm{x} \in \mathcal{D}$ denote the vector for spatial coordinates, $t$ denote time and $u$ denote the true water level depths. Noisy observations of water levels, $\big\{\big(\bm{x}_i, t_j, u_{obs}(\bm{x}_i, t_j)\big); i \in \{1,\ldots, m\}, j \in \{1,\ldots,n_i\}\big\}$ are available at $m$ discrete water well locations, each with an irregular number of $n_i$ samples across time. In this paper, the variables of interest are long-term and seasonal trends of water level fluctuations. In section \ref{Application}, we model water levels $u(\bm{x},t)$ at any $\bm{x}$ by combining linear and sinusoidal time-series models quantifying the long-term and seasonal water level fluctuations respectively. Note that we make the simplifying assumption that water levels vary only across 2D space $\bm{x}$ as discussed in sections \ref{prediction_variables} and \ref{features}. The prediction variables consist of the temporal model parameters, i.e., intercept and slope parameters of the linear model, and amplitude and phase parameters of the sinusoidal model at all $\bm{x} \in \mathcal{D}$. The preceding four temporal model parameters at each spatial location are denoted as $\bm{y} \in \mathbb{R}^d; d=4$. The formulation we present below thus applies to multivariate prediction variables. Given observations of $u$ at wells, noisy estimates $\{\bm{y}(\bm{x}_{i}); i=1,\ldots,m\}$ of the prediction variables may be obtained by linear regression at well locations as discussed in section \ref{prediction_variables}. The goal is to recover the true underlying signal of the prediction variables at any $\bm{x}$. 

The proposed approach of this paper is to model the prediction variables as multivariate Gaussian random processes. A random process is defined as an indexed collection of random vectors. A specific advantage of the GP formulation is that given a set of irregularly sampled observations, the posterior predictive distribution at any query index may be derived analytically. In many cases, prediction variables $\bm{y}$ wont be normally distributed. A normal score histogram transformation (see section \ref{prediction_variables}) may be applied in that case and the GP formulation is specified on the normally transformed variables.

\subsection{Spatial GP regression and limitations}
Consider the baseline case when the GP $a(\bm{x}):\mathbb{R}^2 \rightarrow \mathbb{R}^d$ is indexed over geospatial coordinates $\bm{x}$. Specifically,  
    \begin{linenomath*}
    \begin{equation}
    \label{eq:vanillaGPPrior}
    a(\bm{x}) \sim \mathcal{GP}(0,k(\bm{x},\bm{x}')),
    \end{equation}
    \end{linenomath*}
with zero mean and covariance kernel $k:\mathbb{R}^2 \times \mathbb{R}^2  \rightarrow \mathbb{R}^{d \times d}$. The covariance kernel $k(\bm{x},\bm{x}')$ encapsulates the prior assumptions on the spatial heterogeneity of the random process by quantifying the similarity between two input locations $\bm{x}$ and $\bm{x}'$. Given the above a-priori Gaussian assumptions and well observations, the posterior predictive distribution at any query location $\bm{x}_*$ may be derived by multivariate Gaussian process (GP) regression, also known as cokriging. Cokriging is a very mature methodology for spatial interpolation and regression (see references in section \ref{introduction}). Several variants of the basic kriging approach have been proposed such as simple cokriging, ordinary cokriging and universal cokriging distinguished by how the mean of the GP is specified. In this paper, we work with the simple cokriging approach in which the mean is specified to be a constant across $\mathcal{D}$ as shown in equation \ref{eq:vanillaGPPrior}. 

The main limitations associated with the cokriging approach are as follows:
\begin{enumerate}
    \item  In typical geological settings, repeated measurements of prediction variables at two distinct locations $\{\bm{y}(\bm{x}),\bm{y}(\bm{x}')\}$ are seldom available to make inferences about the form of $k(\bm{x},\bm{x}')$. Therefore, a decision of spatial stationarity over $\mathcal{D}$ is made by assuming $k(\bm{x},\bm{x}')$ depends only on the distances between the input locations as $k(\|\bm{x}-\bm{x}'\|_2)$, where, $\|.\|_2$ denotes $\ell_2$ norm (\cite{goovaerts97}). Note that under the above assumptions, the covariance kernel will be invariant to spatial translations. Thus, any two locations separated by identical distances will be assigned identical covariances. Since hydrological and hydrogeological data commonly exhibit significantly varying spatial correlation scales across the hydrological basin, stationary kernels may lead to overly smooth or rough processes, limiting the predictive ability of the model.
    
    \item As defined later, standard covariance kernels model a range of influence through their length-scale parameters. In other words, posterior predictive uncertainty at a test location beyond the range of influence of well location will be large. Thus, regression over $\mathcal{D}$ will lead to large posterior predictive uncertainty if the observed data is sparsely sampled, which is the case with CV well data. 
\end{enumerate}

A potential solution is to perform the GP regression in an extended feature space $\hat{\bm{x}} = {[\bm{x}, \check{\bm{x}}]}^T, \check{\bm{x}} \in \mathbb{R}^n$, with $T$ being the transpose operator. For modeling groundwater levels, $\check{\bm{x}}$ could pertain to the aquifer hydrogeology, e.g., the depth and lithologic composition of aquifer stratigraphies (see section \ref{features} for context within the CV hydrological basin). The requirement is that $\check{\bm{x}}$ must be known for every $\bm{x} \in \mathcal{D}$ such that random process may be indexed in the combined geospatial and hydrogeological space $\hat{\bm{x}}$ as $a(\hat{\bm{x}})$. Indexing $a(.)$ over $\hat{\bm{x}}$ will address the training data sparsity limitation when testing coordinates have more similarity to the training data coordinates in $\hat{\bm{x}}$-space vs. $\bm{x}$-space. However, note that this does not explicitly bypass the stationarity assumption since the true $a(\hat{\bm{x}})$ may also exhibit non-stationary behavior over $\hat{\bm{x}}$. To handle this, we propose using the GP-DNN formulation which learns to appropriately re-configure the distances between the random process index.   

\subsection{Multivariate regression by hierarchical GP-DNN}
As motivated in section \ref{introduction}, several authors have proposed to handle non-stationary data by modeling a latent space that is able to accommodate the assumptions of stationarity, 
\begin{linenomath*}
\begin{equation}
    \label{eq:hierarchicalGP1}
    \Tilde{\bm{x}}=\phi(\hat{\bm{x}};\bm{\theta}), 
\end{equation}
\end{linenomath*}
where $\Tilde{\bm{x}} \in \mathbb{R}^p$ denotes the latent feature space and $\phi: \mathbb{R}^{n+2} \rightarrow \mathbb{R}^p$ is a feature projection map to be inferred by machine learning. The dimensionality of input feature space is $n+2$ since it also includes the 2-dimensional geospatial coordinates. In the proposed GP-DNN hierarchical model, $\phi(.;\bm{\theta})$ is taken to be a deep neural network with learnable parameters $\bm{\theta}$. The Gaussian random process predictive prior will be indexed over the latent feature space,
    \begin{linenomath*}
    \begin{equation}
    \label{eq:DNNGPPrior}
    a(\Tilde{\bm{x}}) \sim \mathcal{GP}(0, k(\Tilde{\bm{x}},\Tilde{\bm{x}}'))),
    \end{equation}
    \end{linenomath*} 
where $k:\mathbb{R}^p \times \mathbb{R}^p  \rightarrow \mathbb{R}^{d \times d}$ is the covariance kernel. Note that the usage of the term \emph{predictive} indicates probability distributions defined over the prediction variables. We show next how the GP prior predictive distribution may be conditioned to observations to derive the posterior predictive distribution.

\subsubsection{The generative model and conditioning to training data}
In the following, subscripts $\tau$ and $*$ are used to denote training and test data for machine learning. Let $\bm{y}_{\tau}$ denote the $dm\times1$ vector constituting noisy observations of the true signal $a(\Tilde{\bm{x}})$ corresponding to well locations and $\Tilde{X}_\tau$ is the $dm \times p$ latent feature matrix, which is computed from the original $dm \times (n+2)$ input feature matrix $\hat{X}_\tau$ using the DNN. $d$ is the dimensionality of the prediction variable and $m$ is the number of training samples. Since  $a(\Tilde{\bm{x}})$ is assumed to be a GP, its' realizations corresponding to the well locations, denoted by the vector $\bm{a}_{\tau}$, will be distributed according to a multivariate Gaussian distribution. We assume that realizations of $\bm{a}_\tau$ are subsequently corrupted by Gaussian noise yielding the noisy measurements $\bm{y}_{\tau}$. To summarize, the generative model for the well data is specified as,
\begin{linenomath*}
\begin{equation}
    \label{eq:hierarchicalGP2}
    \Tilde{X}_\tau = \phi(\hat{X}_\tau),  \ \bm{a}_\tau|\Tilde{X}_\tau \sim \mathcal{N}(\bm{0},\Tilde{K}_{\tau\tau}), \ \textrm{and}
    \ \bm{y}_\tau|\bm{a}_\tau \sim \mathcal{N}(\bm{a}_\tau,\Sigma_n), 
\end{equation}
\end{linenomath*}
where $\mathcal{N}(.,.)$ denotes the multivariate Gaussian distribution and  $\Tilde{K}_{\tau\tau}$ is the $dm \times dm$ covariance matrix (see section section \ref{mvkernel} for details). Note that operator $\phi(.; \bm{\theta})$, in the first equality above, operates on each row of matrix $\hat{X}_\tau$ and we have abused notation for brevity. The covariance matrix for the noise distribution is specified as $\Sigma_n=diag([\bm{\sigma}_{n_1},\ldots,\bm{\sigma}_{n_d}]^T)$, where $\bm{\sigma}_{n_i}=[\sigma_{n_i}^2,\ldots,\sigma_{n_i}^2]^T$ is a $m\times1$ vector specifying identical observational noise variance $\sigma_{n_i}^2$ for each prediction variable $a(\bm{x})_i$. The covariance kernel parameters, noise levels and DNN architectural variables are treated as hyper-parameters of the model, to be tuned by cross-validation as described in section \ref{training} (see Table \ref{table:hyperParamGPextended} for a complete list of hyper-parameters). Parameters $\bm{\theta}$ of the DNN are trained as discussed in section \ref{gpdnntraining}.

Let $\bm{a}_*$ denote the realization of $a(\hat{\bm{x}})$ at test location $\hat{\bm{x}}_*$. The posterior predictive distribution of $\bm{a}_*$ is obtained by conditioning the prior predictive distribution to the noisy observations. The joint prior distribution of $\bm{y}_\tau$ and $\bm{a}_*$, given the training and test locations, may be specified as
\begin{linenomath*}
\begin{equation}
    \label{eq:jointTrainTest}    
    \begin{bmatrix} \bm{y}_\tau \\ \bm{a}_* \end{bmatrix} | \hat{X}_\tau,\hat{\bm{x}}_* \sim 
    \mathcal{N}\bigg(\bm{0},
    \begin{bmatrix}
    \Tilde{K}_{\tau\tau}+\Sigma_n & \Tilde{K}_{\tau*}\\
    \Tilde{K}_{*\tau} & \Tilde{K}_{**})
    \end{bmatrix}
    \bigg)
\end{equation}
\end{linenomath*}
(\cite{rasmussen06}), where, $\Tilde{K}_{\tau\tau}$ is the covariance matrix for the training locations, $\Tilde{K}_{*\tau}$ and $\Tilde{K}_{\tau*}$ contain the covariances between the training and test locations, while $\Tilde{K}_{**}$ is the covariance matrix for test locations. Since the joint distribution of $\bm{y}_\tau$ and $\bm{a}_*$ is a Gaussian distribution, the distribution of $\bm{a}_*$, conditioned on the training observations and the training and test features, is also a Gaussian distribution derived as
\begin{linenomath*}
\begin{equation}
    \label{eq:cokriging}
    \bm{a}_*|\bm{y}_\tau,\hat{X}_\tau,\hat{\bm{x}}_* \sim \mathcal{N}\bigg(\Tilde{K}_{*\tau}\big[\Tilde{K}_{\tau\tau}+\Sigma_n\big]^{-1}\bm{y}_\tau,
    \Tilde{K}_{**}-\Tilde{K}_{*\tau}\big[\Tilde{K}_{\tau\tau}+\Sigma_n\big]^{-1}\Tilde{K}_{\tau*}\bigg).
\end{equation}
\end{linenomath*}
Here, $\Tilde{K}_{*\tau}\big[\Tilde{K}_{\tau\tau}+\Sigma_n\big]^{-1}\bm{y}_\tau$ is the cokriging estimate of the posterior predictive mean and $\Tilde{K}_{**}-\Tilde{K}_{*\tau}\big[\Tilde{K}_{\tau\tau}+\Sigma_n\big]^{-1}\Tilde{K}_{\tau*}$ is the posterior predictive covariance.

\subsubsection{The multivariate kernel} \label{mvkernel}
The assumption of covariance kernel stationarity is made in the latent space by taking
\begin{linenomath*}
\begin{equation}
\label{eq:stationaryKernel}
k(\Tilde{\bm{x}},\Tilde{\bm{x}}') = k(\|\Tilde{\bm{x}}-\Tilde{\bm{x}}'\|_2), \forall \Tilde{\bm{x}},\Tilde{\bm{x}}'.
\end{equation}
\end{linenomath*}
By definition, the covariance kernel should be a symmetric, positive definite function (\cite{paciorek03}). Several valid covariance functions have been studied in the geostatistical and ML literature such as the squared exponential and the Matérn kernel. To ensure kernel validity in the multivariate regression setting, we employ the linear model of coregionalization (\cite{journel78}, \cite{deiaco03}) which models all the components of the multivariate process as linear combinations of the same underlying permissible random processes. This states that the kernel, when specified as
\begin{linenomath*}
\begin{equation}
\label{eq:covKernel}
    k(\bm{x},\bm{x}')= \sum_i K_{amp}^i k_{valid}^i(\|\Tilde{\bm{x}}-\Tilde{\bm{x}}'\|_2),
\end{equation}
\end{linenomath*}
will be a valid positive semi-definite kernel if $K_{amp}^i$ is a $d \times d$ positive semi-definite matrix and $k_{valid}^i: \mathbb{R}^p\times\mathbb{R}^p \rightarrow \mathbb{R}$ is a permissible positive semi-definite kernel for each $i$. $K_{amp}^i$ contains the variance and covariance scaling factors for the prediction variables. If each component of $a(.)$ is normalized to have unit variance, then the diagonal elements of $K_{amp}^i$ will have unit magnitude and off-diagonal elements specify the correlation coefficent for each variable pair. In this paper, we choose $i=1$ and  $k_{valid}$ to be the Matérn kernel $k_{Mat\Acute{e}rn}^{\nu=2.5}:\mathbb{R}^2 \times \mathbb{R}^2  \rightarrow \mathbb{R}$, where $\nu$ is the Matérn parameter controlling the roughness of the kernel. The Matérn kernel allows greater control on the roughness of the random process through the $\nu$ parameter (\cite{stein99}) and $\nu=2.5$ is a common choice for machine learning applications (\cite{rasmussen06}). \ref{Appendix: Covariance} contains additional details on how the covariance matrices are constructed from the Matérn kernel.

\subsubsection{GP-DNN end-to-end training} \label{gpdnntraining}
DNN parameters $\bm{\theta}$ will be trained end-to-end along with the GP layer by maximization of the data likelihood distribution as described below. The dimensionality $p$ of the latent feature space $\Tilde{\bm{x}}$ and the DNN architectural parameters, such as the number of hidden layers and the number of neurons in each hidden layer are treated as hyper-parameters, to be tuned by cross-validation. From equation \ref{eq:hierarchicalGP2}, it follows that the training data likelihood
\begin{linenomath*}
\begin{equation}
    \bm{y}_\tau|\hat{X}_\tau; \bm{\theta} \sim \mathcal{N}(\bm{0},\Tilde{K}_{\tau\tau}+\Sigma_n).
\end{equation}
\end{linenomath*}
Since the Gaussian likelihood has an analytical expression, parameters $\bm{\theta}$ may be estimated by minimizing the negative of log-likelihood
\begin{linenomath*}
\begin{equation}
    \label{eq:logLikelihood}
    \textrm{log} \ p(\bm{y}_\tau|\hat{X}_\tau; \bm{\theta})=-\frac{1}{2}\bm{y}_\tau^T\big[\Tilde{K}_{\tau\tau}+\Sigma_n\big]^{-1}\bm{y}_\tau-\frac{1}{2}\textrm{log} \ |\Tilde{K}_{\tau\tau}+\Sigma_n|+\textrm{constant}
\end{equation}
\end{linenomath*}
(\cite{rasmussen06}, \cite{calandra2016}, \cite{wilson2016}, \cite{bradshaw2017}). Taking derivatives of $\textrm{log} \ p(\bm{y}_\tau|\hat{X}_\tau)$ w.r.t the parameter $\theta_k$, we obtain
\begin{linenomath*}
\begin{equation}
    \label{eq:Derivative}
    \begin{gathered}
    \frac{\partial}{\partial \theta_k} p(\bm{y}_\tau|\hat{X}_\tau;\bm{\theta}) =
    \frac{1}{2}\bm{y}_\tau^T\big[\Tilde{K}_{\tau\tau}+\Sigma_n\big]^{-1} \frac{\partial \Tilde{K}_{\tau\tau}}{\partial \theta_k} \big[\Tilde{K}_{\tau\tau}+\Sigma_n\big]^{-1} \bm{y}_\tau \\
    - \frac{1}{2}tr\big(\big[\Tilde{K}_{\tau\tau}+\Sigma_n\big]^{-1} \frac{\partial \Tilde{K}_{\tau\tau}}{\partial \theta_k}\big).
    \end{gathered}
\end{equation}
\end{linenomath*}

In the equality above, we used identities for derivative of inverse matrix , $\frac{\partial K^{-1}}{\partial \theta} = -K^{-1}\frac{\partial K}{\partial \theta} K^{-1}$, and the derivative of matrix log determinant $\frac{\partial \textrm{log}|K|}{\partial \theta} = tr(K^{-1}\frac{\partial K}{\partial \theta})$. Given that the $k_{Mat\Acute{e}rn}^{\nu=2.5}(.)$ is a differentiable function w.r.t to its inputs, the entries of $\frac{\partial \Tilde{K}_{\tau\tau}}{\partial \theta_k}$ may be analytically derived by the back-propagation algorithm \cite{bishop06}. Parameters $\bm{\theta}$ may then be trained by typical stochastic gradient descent algorithms commonly employed to train deep learning networks (Algorithm \ref{alg:GPextendedTrain}). Additional details on training and hyper-parameter tuning are provided in Appendix \ref{Appendix: GP-DNN training}.

\subsubsection{Cross-validation statistics for GP based regression} \label{crossvalidation}
The generalization power of GP regression models will be assessed with a test set constituting of randomly sampled well locations that will be held out and kept blind during model training and hyper-parameter tuning. These wells will be referred to as blind wells in our study. Following is a discussion of two specific cross-validation statistics evaluated on the test set, (1) likelihood under the posterior predictive distribution, and (2) deviation of the chi-square plot from the identity function, that are used to compare the GP regression model performance in section \ref{training}.

\paragraph{Posterior predictive likelihood} \label{likelihood_interpretation}
Let $\hat{X}_*$ be the $dm_* \times (n+2)$ be the feature matrix and $\bm{y}_*$ be the $dm_* \times 1$ vector containing observations of the prediction variables computed from the test set. Following from equation \ref{eq:cokriging}, the negative log-likelihood of the Gaussian posterior predictive distribution on test set prediction variables given the training set and test feature matrix
\begin{linenomath*}
\begin{equation}
    \label{eq:likelihood_metric}
    \begin{gathered}
        -\log p(\bm{y}_*|\hat{X}_\tau,\bm{y}_\tau, \hat{X}_*)=\frac{1}{2}\big[\bm{y}_*-\bm{\mu}\big]^TK^{-1}\big[\bm{y}_*-\bm{\mu}\big]+\frac{1}{2}\log |K|+\frac{dm_*}{2}\log 2\pi,
    \end{gathered}
\end{equation}    
\end{linenomath*}
where, $\bm{\mu}=\Tilde{K}_{*\tau}\big[\Tilde{K}_{\tau\tau}+\Sigma_n\big]^{-1}\bm{y}_\tau$ and $K=\Tilde{K}_{**}-\Tilde{K}_{*\tau}\big[\Tilde{K}_{\tau\tau}+\Sigma_n\big]^{-1}\Tilde{K}_{\tau*}$ are the cokriging estimates of the posterior predictive mean and covariance. The first term in the R.H.S. of equation \ref{eq:likelihood_metric} may be interpreted as half of the squared Mahalanobis distance (\cite{mahalanobis36})
\begin{linenomath*}
\begin{equation}
    \label{eq:mahalanobis}
    \mathcal{M}_D^2(\bm{y}_*;\bm{\mu},K) = \big[\bm{y}_*-\bm{\mu}\big]^TK^{-1}\big[\bm{y}_*-\bm{\mu}\big].
\end{equation}    
\end{linenomath*}
Given mean $\bm{\mu}$ and covariance matrix $K$, Mahalanobis distance computes the statistically standardized distance of $\bm{y}_*$ from the mean. The inverse covariance matrix $K^{-1}$ serves the purpose of standardizing each coordinate of $\mathbb{R}^{dm_*}$ by corresponding variance and removing inter-coordinate correlations as estimated by the posterior predictive distribution (\cite{etherington19}). The Mahalanobis distance thus acts as an indicator of the fit of the test data under the estimated posterior predictive distribution. The second and third terms in the R.H.S. of equation \ref{eq:likelihood_metric} relate to the normalization constant of the multivariate Gaussian probability density and consequently the volume under the multivariate density function. For a given dimensionality of the output space, $\frac{1}{2}\log |K|$ could be interpreted in terms of the predictive model complexity. $|K|$ is the generalization of variance in multivariate settings (\cite{wilks32}), and thus a larger value for $|K|$ implies a more complex model that will be able to explain larger variability in the data. In other words, rewarding lower values of the negative log-likelihood statistic encourages simpler models that fit the data better.

\paragraph{Chi-square quantile-quantile plots for GP} 
Going beyond the data likelihood, we propose to employ quantile-quantile (Q-Q) plots to assess the goodness of fit of the predicted uncertainty intervals to the test data. The Q-Q plot is a standard statistical tool for evaluating whether empirical data belong to a specified theoretical probability distribution through a graphical comparison of the empirical quantiles to their theoretical counterparts (\cite{gnanadesikan68}). While it is generally difficult to generalize Q-Q plots to multivariate data (\cite{easton90}), multivariate normality can be tested using the chi-square plot (\cite[][see section 4.6]{johnson07}). We show how the chi-square plot approach may be extended to multivariate GPs. For notational simplicity, we present the results considering spatial GP regression but it is straightforward to extend the treatment to GP-DNN. Consider a randomly selected set of feature coordinates $\{\bm{x}_{1}, \bm{x}_2, \ldots, \bm{x}_{m_*}\}$ from $\mathcal{D}$ where noisy observations $\{\bm{y}_1, \bm{y}_2, \ldots, \bm{y}_{m_*}\}$ of $d$-variate prediction variables are available. The objective is to determine the accuracy of the hypothesis that each observation $\bm{y}_i$ is a sample from the corresponding predictive posterior distribution in the set $$\{\mathcal{N}(\bm{\mu}_1,K_1), \mathcal{N}(\bm{\mu}_2,K_2), \ldots, \mathcal{N}(\bm{\mu}_{m*},K_{m*})\},$$ where $\mathcal{N}(\bm{\mu}_i,K_i)$ is distribution of $a(\bm{x}_i)|X_\tau,\bm{y}_\tau,\bm{x}_i$ as estimated using equation \ref{eq:cokriging} $\forall \bm{x}_i \in \{\bm{x}_1, \bm{x}_2,\ldots,\bm{x}_{m_*}\}$. As proved by \citet[][see result 4.7]{johnson07}, 
\begin{linenomath*}
    \begin{equation}
        \bm{y}\sim \mathcal{N}(\bm{\mu},K) \implies \mathcal{M}_D^2(\bm{y}; \bm{\mu}, K) \sim \mathcal{X}_d^2,
    \end{equation}
\end{linenomath*}
where $\mathcal{X}_d^2$ is the chi-square distribution with $d$ degrees of freedom with $\mathcal{M}_D^2(.)$ being estimated by equation \ref{eq:mahalanobis}. It directly follows from this result that if the test locations were sampled independently and the GP regression robustly estimated the associated posterior predictive means and covariance matrices, the set $$\{\mathcal{M}_D^2(\bm{y}_1; \bm{\mu}_1, K_1)\}, \mathcal{M}_D^2(\bm{y}_2; \bm{\mu}_2, K_2), \ldots, \mathcal{M}_D^2(\bm{y}_{m_*}; \bm{\mu}_{m_*}, K_{m_*})\}$$ will be expected to contain roughly independent samples of $\mathcal{X}_d^2$. The chi-square Q-Q plot is a scatter plot of the empirical quantiles of the standardized Mahalanobis distances and theoretical quantiles of the chi-square distribution. The quantiles and corresponding cumulative probability values of the empirical data distribution are derived from the ordered set of sample distances $$\{\mathcal{M}_D^2(\bm{y}_i; \bm{\mu}_i, K_i)_{(1)}\le \mathcal{M}_D^2(\bm{y}_j; \bm{\mu}, K_j)_{(2)} \le \ldots \le \mathcal{M}_D^2(\bm{y}_k; \bm{\mu}_k, K_k)_{(m_*)} \}.$$ The cumulative probability of the empirical quantile is subsequently mapped to the corresponding theoretical quantile of $\mathcal{X}_d^2$. If the empirical data distribution is representative of the theoretical distribution, within effects of limited sample availability at all test locations and theoretical simplifications of the real data noise, the Q-Q pairs will roughly plot along the identity line. Deviations of the Q-Q scatter trend from the identity line can thus be used for cross-validation of the posterior predictive uncertainty estimates. 

\section{Application to Central Valley (CV)} \label{Application}
\begin{figure}[!ht]
    \centering
    \includegraphics[scale=0.65]{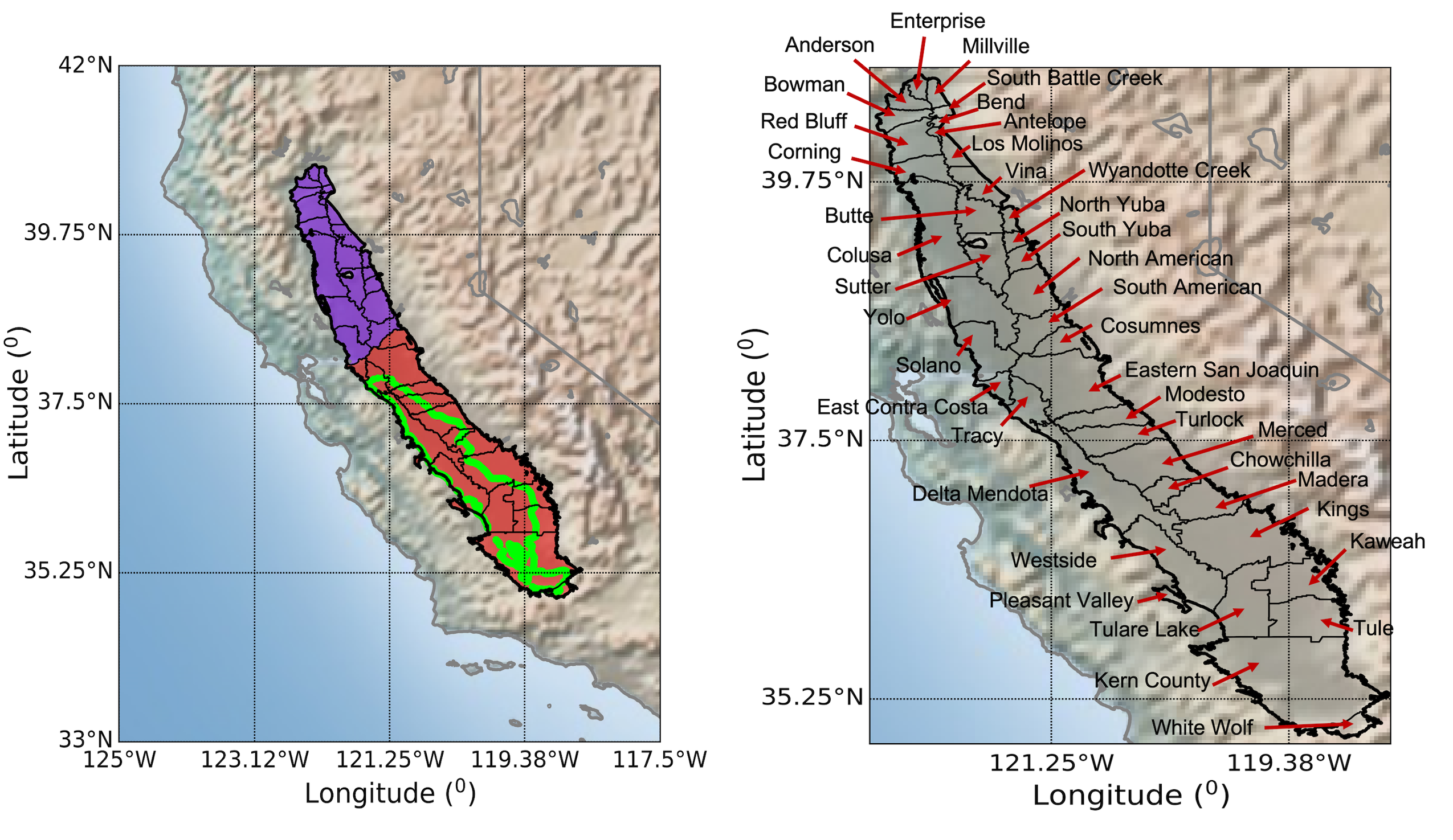}
    \caption{Extent of CV outlined in black with groundwater subbasins. (Left) Sacramento and San Joaqin Valley areas have been colored by purple and red. Mapped extent of the Corcoran clay is shown in lime green. (Right) Nomenclature of the CV's groundwater subbasins.}
    \label{fig:cv_map}
\end{figure}

Our study area is Central Valley covering approximately 20,000 square miles in central California (Figure \ref{fig:cv_map}). Groundwater pumping is a primary source of water support for its vast agricultural system (\cite{williamson89}, \cite{bertoldi91}, \cite{faunt09}). For convenience of discussion, the study area is typically divided into the northern Sacramento valley (SV) and southern San Joaquin valley (SJV). The CV groundwater basin has been delineated into several subbasins based on factors such as hydrogeologic barriers or institutional boundaries (\cite{dwr03}), which have been overlain on the CV map shown in Figure \ref{fig:cv_map}. The sediments of the underlying aquifer system were derived from the surrounding Sierra Nevada and the Coast Ranges. Defining stratigraphic units in the CV aquifer system has generally been difficult due to absence of distinct lithologic changes (\cite{faunt09}). The SV sediments have been determined to constitute of coarse-grained alluvial sediments interbedded with localized fine-grained sediments attributed to low-energy drainage basins. In the SJV, the hydrogeologic makeup is described in terms of an upper semi-confined and lower confined aquifer zone, separated by a confining unit. Three intermixing hydrogeologic units, namely Coast Ranges alluvium, Sierran alluvial deposits, and flood-basin deposits, form the constituents of the upper semi-confined aquifer zone (\cite{laudon91}). Fine-grained alluvium, predominantly derived from the Coast Ranges, are present in the form of spatially discontinuous lenticular shapes, comprising approximately half of the volumetric fill. In contrast to the SV, within the SJV there is a distinct and spatially continuous confining unit dividing the upper and lower aquifers consisting of low-permeability clay deposits known as the Corcoran Clay. The spatial extent of the Corcoran clay is well-mapped (Figure \ref{fig:cv_map}) as described in section \ref{data}. 

While the general hydrogeologic characteristics of the CV has been well-studied, physically modeling groundwater flow in the CV encounters large uncertainty resulting  from the regional and local stratigraphic variations of the hydraulic and storage properties required in flow modeling studies. We undertake a machine learning approach to address this spatial uncertainty. Specifically, our approach involves the Gaussian process methodology formulated in section \ref{methodology} to model groundwater level long-term and seasonal variability trends using hydrogeologic and surface deformation features.

\subsection{Available data} \label{data}
We describe below the well and lithological texture datasets available to us in the study area. Note that while the proposed methodology will work on irregular, non-discretized data, we perform discretization of the well data for ease of analysis with the lithologic texture data which is available in gridded manner. We choose a 1 square mile spatial resolution for every cell in the modeling grid over CV to correspond with that of the sediment texture data.

\begin{figure}[!ht]
    \centering
    \includegraphics[scale=0.42]{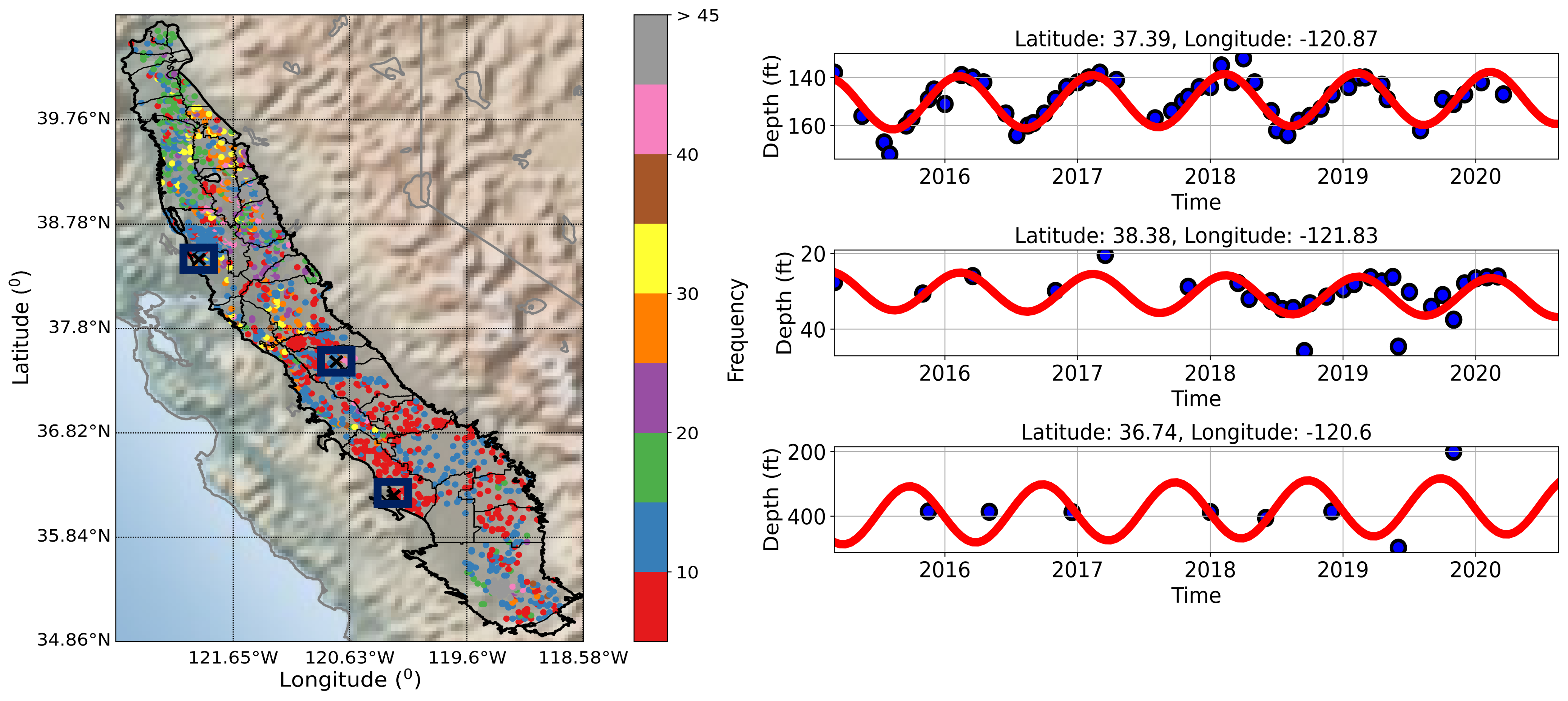}
    \caption{(Left) Well locations colored by the frequency of data samples. Wells shown in the right plot are highlighted with black crosses. (Right) Water levels measurements at three wells are shown in blue circles. Best fitting long term and seasonal trend line, estimated using equation \ref{eq:linearReg1}, is overlain in red on well data.}
    \label{fig:data_density}
\end{figure}

\begin{enumerate}
    \item \emph{Water level data}: We use water level time series data across the CV as compiled and processed  by \cite{kim21} using groundwater well datasets obtained from the California Department of Water Resources (DWR) and United States Geological Survey (USGS). The dataset consists of measurements from approximately 4500 wells across our study area. Well screen depth information is not available. We discarded approximately half of the wells for having too few samples (less than 8) along time axis. A few wells were ignored as they clearly contained outlier samples. The well data were spatially aggregated into the modeling grid over the CV. Many grid cells, post aggregation, contained multiple individual wells, which should not be co-mingled within the time series. Additionally, there were cases where two different time-series were presented from two different data sources for one co-located grid cell, leading to confusion which measurement was to be considered for this study. To work around this issue and to apply a uniform standard across the modeling grid of study, the most temporally robust well record within each grid cell was selected. After spatial aggregation at the modeling grid resolution, data is available at approximately 1750 spatial locations (Figure \ref{fig:data_density}). Temporally, the data was available till August 2020 and we considered a rough five year period starting from March 2015 for this study. The well time series data was averaged at biweekly intervals. This resulted in well data aggregated into the spatio-temporal grid having 400, 220 and 132 cells along latitude, longitude and time axes respectively. Figure \ref{fig:data_density} shows the time series data at three wells. The footprint of the dry-wet seasonal cycle on the water levels can be clearly observed in the top plot. The best fitting long-term and seasonal trend (red line) to the well data is discussed further in section \ref{prediction_variables}.
    
    \item \emph{Lithological texture data}: Information on subsurface hydrogeology in the form of volumetric proportion of coarse and fine grained sediments (commonly referred to as lithological texture) is available on a uniformly discretized grid across the CV, as modeled in previous work by \citet{cvhm22}. The authors generated the texture model by 3D kriging of texture observations derived from approximately 8500 drillers logs. To address the challenges of non-stationarity in kriging, the authors divided the study area areally and vertically into several modeling domains. Kriging was performed separately across each groundwater subbasin (see Figure \ref{fig:cv_map}) and aggregated vertically into 13 layers. The stratigraphy of the 3D model is largely determined by the structure of the Corcoran Clay, represented by layers 6-8 in the model. The remainder of the layers are divided between the upper semi-confined and lower confined aquifers. 
\end{enumerate}

\subsection{Training dataset generation}
The objective is to estimate the posterior predictive distribution $\bm{a}_*|\bm{y}_\tau,\hat{X}_\tau,\hat{\bm{x}}_*$ as specified in equation \ref{eq:cokriging} for all $\bm{x}_*$, in the CV grid. We randomly distributed available wells into training, validation and test sets of sizes 1550, 100 and 100 wells respectively. We also consider a robust test set of 90 wells, which is created by removing 10 outlier wells from the original test set using the outlier detection scheme described in section \ref{training}. For each evaluation set, we created target variables and input features as discussed below.

\subsubsection{Prediction variables} \label{prediction_variables}
We seek to predict quantitative metrics of groundwater level fluctuation during the 2015-2020 study period. As discussed in the section \ref{introduction}, it is useful to model the seasonal and long-term water level fluctuation trends over time. In this paper, we assume that measured CV water level depths vary in 2D only as a function of spatial latitude and longitude coordinates $\bm{x}$ and time $t$. Additionally, we assume that water level time series data from 2015-2020 at each $\bm{x}$ may be decomposed as a linear model for the long-term signal and a sinusoidal model for the seasonal signal. Mathematically, water level at $\bm{x}$ varies through $t$ as
\begin{linenomath*}
    \begin{equation}
    \label{eq:linearReg1}
        \begin{gathered}
        u(\bm{x},t)=a_1(\bm{x})+a_2(\bm{x})t+
        a_3(\bm{x})sin\bigg( \frac{2\pi t}{\lambda}+a_4(\bm{x}) \bigg)\\
        \end{gathered}
    \end{equation}
\end{linenomath*}
where, $a_1(\bm{x})$, $a_2(\bm{x})$, $a_3(\bm{x})$ and $a_4(\bm{x})$ denote the intercept, slope, amplitude and phase parameters respectively. Note that the simplifying assumption of 2D variability of water levels is made to facilitate initial evaluation of the efficacy of GP methodological development conducted in this paper for hydrological modeling. Extending proposed methodology for rigorous 3D-modeling of groundwater levels, where vertical connectivities between all 13 aquifer model layers are accounted for, is left as future work. We also chose to make the simplifying assumption of a single long-term and seasonal model across 2015-2020 as the well data is very sparsely sampled along time at several well locations (see well data frequency in Figure \ref{fig:data_density}) and complex models may overfit to the data. However, note that the proposed methods can easily be extended to other complicated temporal models, for instance the B-spline integrated with multi-period sinusoidal model considered by \cite{riel2018}. 

\begin{figure}[!ht]
    \centering
    \includegraphics[scale=1.55]{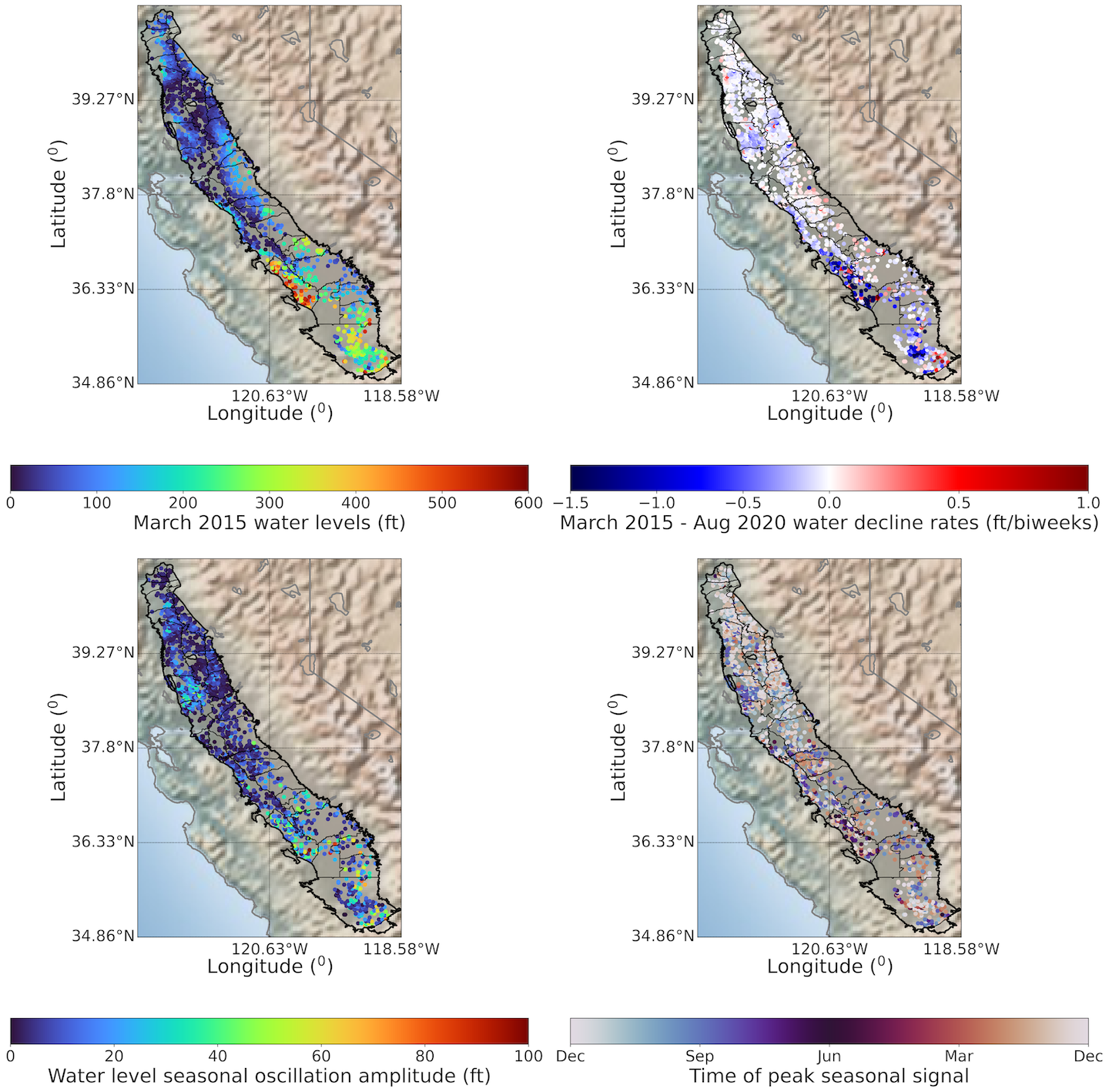}
    \caption{Water level long-term and seasonal model parameters fitted by linear regression on well time series data.}
    \label{fig:targets_well}
\end{figure}

The long-term and seasonal parameter fields are estimated independently at each well by solving a linear regression problem with the corresponding time series data (see \ref{Appendix: Time series regression}). Figure \ref{fig:data_density} shows the modeled water level signal along with observed data at two well locations, while Figure \ref{fig:targets_well} shows the estimated parameters at all wells. The results indicate that in March 2015 (start of our analysis period), water levels were relatively deeper in the SJV as compared to the SV. During the next five years, wells exhibit both long-term decline and uplifts of water levels, with largest uplifts observed in the wells of the Westside subbasin and Kern county. The seasonal amplitude signal generally has high magnitudes in the southern SJV. A common feature across the long-term and seasonal trend parameters is the smoother variabiilty in the northern two-thirds of the valley, with greater spatial heterogeneity in the southern San Joaquin basin, likely a manifestation of the underlying hydrogeologic heterogeneity as discussed in section \ref{Results}.  Another complicating factor is data sparsity since most wells in the southern SJV contain $<15$ samples (Figure \ref{fig:data_density}). Thus, the linear regression trend estimates are expected to be noisier. The GP methodology accounts for this noise through the estimated noise-level matrix $\Sigma_n$ (see equation \ref{eq:hierarchicalGP2}). As discussed in section \ref{Results}, the hierarchical GP-DNN model correctly identifies this data uncertainty by predicting wider uncertainty intervals in the southern SJV.

It should be noted that trend data at wells cannot be expected to be Gaussian in nature, for instance, water levels depths cannot assume negative magnitudes and are skewed towards positive values. Thus, the GP methodology is not directly amenable to the water level data. A simple yet effective approach to handle this issue is to perform a normal score transform (\cite{journel78}), which transforms the sample data histogram of each prediction variable into the standard Gaussian distribution. In this paper, the GP regression is performed on the transformed normal variables and the regression outputs, as shown in section \ref{Results}, were obtained by subsequent back-transformation to replicate the original well data sample histogram. 

\subsubsection{Feature variables} \label{features}
\begin{figure}[!ht]
    \centering
    \includegraphics[scale=0.6]{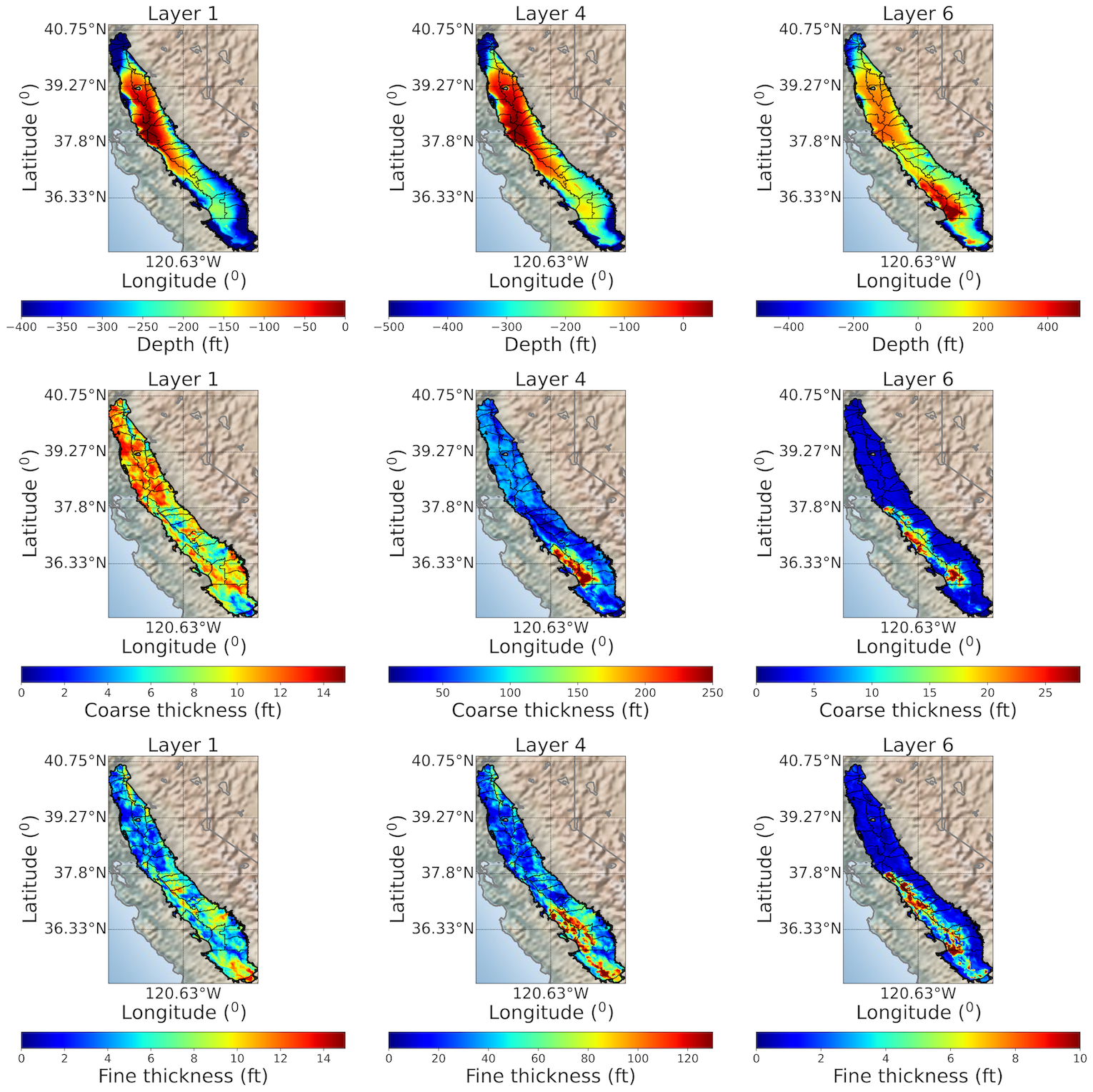}
    \caption{Depth to layer tops (top row), coarse-grained sediment thickness (middle row) and fine-grained sediment thickness (bottom row) for three different texture model layers. Top of layer 6 corresponds to top of Corcoran clay.}
    \label{fig:features_texture}
\end{figure}

We consider the following features as input to the GP regression as discussed below. 
\begin{enumerate}
    \item \emph{Geospatial coordinates $\bm{x}$}: The baseline features we consider are the latitude and longitude coordinates. The baseline GP regression employs only $\bm{x}$ as features as discussed earlier, while GP regression in extended feature space uses $\bm{x}$ and additional features as discussed below.
    \item \emph{Hydrogeological features $\check{\bm{x}}$}: Amongst the several factors that groundwater flow depends on, geologic variability of the underlying aquifer plays a crucial role. Hydraulic head evolves in 3D inside an aquifer and can be physically described through the 3D groundwater flow equation (\cite{harbaugh05}),
    \begin{linenomath*}
    \begin{equation}
        \label{eq:diffusion3D}
        \frac{\partial}{\partial x}(K_{xx}\frac{\partial h}{\partial x}) + \frac{\partial}{\partial y}(K_{yy}\frac{\partial h}{\partial y}) + \frac{\partial}{\partial z}(K_{zz}\frac{\partial h}{\partial z}) + W =S_s\frac{\partial h}{\partial t},
    \end{equation}
    \end{linenomath*}
    where $h$ is the hydraulic head, $x, y, z$ denote the spatial dimensions, $t$ denotes time, $S_s$ is the specific storage coefficient, $K_{xx}, K_{yy}, K_{zz}$ represent hydraulic conductivity along the spatial dimensions and $W$ represents the flow source/sink term. For unconfined aqyuifers, specific yield is used as the storage coefficient. While the general groundwater flow equation presented above calculate hydraulic head in all three spatial directions, we use a 2D assumption for groundwater level variation (see section \ref{prediction_variables}). For subsurface rocks, the hydraulic and storage properties will vary depending on several factors, including the lithologies, lithological composition and microstructure (structural arrangement of the sediments and pores) of the porous rock medium (\cite{mavko09}). In general, these properties may be measured by field well tests or laboratory tests on rock core samples. For instance, hydraulic conductivity for a core sample may be measured by a permeameter as 
    \begin{linenomath*}
    \begin{equation}
        \label{eq:permeability}
        K = \frac{\Delta V}{\Delta t} \frac{Z}{Ah}
    \end{equation}
    \end{linenomath*}
    (\cite{todd05}), where $\Delta V$ represents the volumetric flow of water in time $\Delta t$, $Z$ is the thickness of the sample, $A$ is the area of the sample and $h$ is the hydraulic head. Similarly, specific storage coefficient is defined as the volume of water retained or released from a porous medium per unit volume of the aquifer per unit change in $h$. While corresponding in-situ measurements are irregularly available across the CV, the 3D lithological texture model (section \ref{data}) may be used as a proxy for the aquifer lithology (\cite{faunt09}), hydraulic and storage properties. 
    
     In our methodology, the GP-DNN model will attempt to discover useful correlations between the observed water levels and the texture features. Given the texture model, we assume that the aquifer system rocks consists of two lithological end-members, the coarse-grained and fine-grained lithologies. For each layer of the texture model, the following features related to the depth and thickness of the lithologies are extracted. Thickness features were specifically chosen since effective hydraulic or storage properties across a sediment column will depend on the volumetric proportions of lithological end-members (equation \ref{eq:permeability}).
    \begin{enumerate}
        \item \emph{Coarse-grained sediment thickness}: For the $i^{th}$ layer, the thickness of coarse-grained sediments at location $\bm{x}$ is computed as
        $$
            z_{coarse,i}(\bm{x})=f_{coarse,i}(\bm{x})z_i(\bm{x}), \forall i,
        $$
        where $f_{coarse,i}(\bm{x})$ denotes the volumetric fraction of coarse grained sediments at location $\bm{x}$ and layer $i$ and $z_i(\bm{x})$ denotes the depth thickness of layer $i$. Directly using $f_{coarse,i}$ as a feature may be misleading to the machine learning model since the thickness of the layers may vary significantly across $\bm{x}$. We expect $z_{coarse,i}(\bm{x})$ to be informative on effective volume of the coarse-grained lithology for the layer at any given $\bm{x}$.
        \item \emph{Fine-grained sediment thickness}: Assuming only two-end member lithologies, we compute a similar effective volume feature for the fine-grained lithologies
        $$
            z_{fine,i}(\bm{x})=(1-f_{coarse,i}(\bm{x}))z_i(\bm{x}), \forall i.
        $$
        \item \emph{Depths to layer tops}: We also incorporate the depth to each layer top as a feature. Given that the properties of the layers below the water levels will vary due to fluid presence in the pores, it is desirable that the neural network is able to extract any potential correlations between the layer top depths and the water levels. Note that the depths to layer tops of the texture model were available as measured from the mean sea level. To make the datum equivalent to the water level depths which are measured from the surface and contain effects of surface topography, surface elevations as obtained from the NASA Digital Elevation Model (\cite{nasadem21}) were added to the layer depths at each grid location. 
    \end{enumerate}
    Figure \ref{fig:features_texture} shows the layer top depths, coarse and fine grained sediment thicknesses for three texture model layers. Note that we use a total of 39 hydrogeological features (3 for each of the 13 layers) in our analyses.
\end{enumerate}

\subsection{Training, hyper-parameter tuning and cross-validation with blind wells} \label{training}
The baseline GP model and GP-DNN models are regressed using a normalized training set containing pairs of feature and prediction variables. The input features variables in the training set were normalized to have zero mean and unit variance per feature category, i.e., latitude, longitude, coarse sediment thickness, fine sediment thickness and depths to layer tops. We underscore that for the hydrogeological features a single normalization is applied across all the model layers per feature category to preserve inter-layer feature correlations. The prediction variables were normal score transformed as discussed in section \ref{prediction_variables}, hence also have zero mean and unit variance in the training set.  We briefly summarize the model training and hyper-parameter tuning with detailed description of the results presented in \ref{Appendix: trainingTuning}.  For the baseline case, the  posterior predictive distribution can be derived analytically and requires no explict training. The GP-DNN model contains a DNN in the bottom layer of the hierarchy. The DNN architecture constitutes several hidden neural network layers, each of which consists of a number of neurons with trainable weight and bias parameters $\bm{\theta}$, and a multivariate output layer. By hyper-parameter tuning as described below, the optimal DNN architecture was found to consist of 2 hidden layers with 33 neurons in each layer. The dimension of the output latent space $p$ was tuned to be 12. A standard GP model is finally regressed in the space of DNN outputs $\Tilde{\bm{x}}$. Training of the DNN parameters is performed end-to-end by stochastic gradient descent. 

The GP model requires specification of hyper-parameters related to the anisotropic length-scales of input variables (see section \ref{Appendix: Covariance}), observational noise levels of the target variables and the amplitudes of the kernel in $K_{amp}$. The GP-DNN model additionally requires hyper-parameters related to the DNN architecture, such as  the number of hidden layers, and to the training algorithm. Hyper-parameter tuning is performed by finding the best fit under cross-validation. 3000 sample sets of hyper-parameters are generated by random sampling over the range of variability. The negative log-likelihood cross-validation statistic (see section \ref{crossvalidation}) is evaluated on the validation set using the 3000 hyper-parameter sets, and the one that optimizes the statistic is chosen. Table \ref{table:crossValidationMetric} compares the final log-likelihood statistic for the two models considered across different evaluation sets. Also shown is the root mean square error (RMSE) between the predicted posterior mean and prediction variables $\bm{y}$. The training set is the set of wells used for training of the GP-DNN model parameters, while the validation set is the set of wells used for hyper-parameter tuning. The test set is kept blind completely and used only for final validation. We also consider a robust test set  by removing 10 outlier wells from the test set as described below. Note that the validation statistics are computed on the normalized evaluation sets. Given that the DNN has a 1D architecture and the GP posterior predictive distributions are computed analytically, end-to-end training of the DNN model for each hyper-parameter scenario takes about a minute to complete on a machine having 32GB random-access memory (RAM) with a single 32GB graphics processing unit (GPU ). For effectively tuning larger sized deep learning models, efficient hyper-parameter tuning frameworks based on Bayesian optimization (\cite{akiba19}) may be considered. 

While GP-DNN exhibits slightly higher RMSE in the mean estimates, the GP-DNN regression outperforms the conventional GP regression in terms of the uncertainty estimates as evidenced by the higher likelihood of the well data under the estimated posterior predictive distribution. In Table \ref{table:crossValidationMetric}, we have decomposed the negative log-likelihood values according to the R.H.S. of equation \ref{eq:likelihood_metric} to aid interpretability. As expected, the baseline GP model offers simpler models and performs better in terms of model complexity as evidenced by the relatively lower $\frac{1}{2} \log |K|$ values. However, the GP-DNN model performs significantly better in explaining the data variability as demonstrated by the significantly lower standardized Mahalanobis distances of the evaluation set from the predicted probability density. We underscore that the objective of the paper is not to find the best mean model, rather derive an informative predictive posterior that is able to robustly quantify the uncertainty due to the real data irregularity and noise. We use chi-square Q-Q plots next to further bolster the claim that this is achieved with the GP-DNN posterior estimates.

\begin{table}[!ht]
\begin{center}
\begin{tabular}{ |c|c|c|c|c| }
\hline
    \multirow{2}{*}{\textbf{Evaluation set}} & \multicolumn{2}{|c|}{\textbf{Baseline GP}} & \multicolumn{2}{|c|}{\textbf{GP-DNN}} \\
    \cline{2-5}
    & \makecell{posterior predictive \\ $-$ log-likelihood \\ ($ \frac{1}{2}\mathcal{M}_D^2+\frac{1}{2}\log |K|+ $ \\$ \frac{dm_*}{2}\log 2\pi$)} & RMSE & \makecell{posterior predictive \\ $-$ log-likelihood\\ ($ \frac{1}{2}\mathcal{M}_D^2+\frac{1}{2}\log |K|+ $ \\$ \frac{dm_*}{2}\log 2\pi$)} & RMSE\\
    \hline
    \makecell{Training set \\ (1550 wells)} & \makecell{5211292.05 \\ (5222822.08 - 17205.39 + \\ 5675.36)} & 0.64 & \makecell{12217.34 \\ (12009.6 - 5467.62 + \\ 5675.36)} & 0.45\\
    \hline
    \makecell{Validation set \\ (100 wells)} & \makecell{2784.99 \\ (3017.2 - 599.79 + \\ 367.58)} & 0.77 & \makecell{446.51 \\ (256.65 - 177.72 + \\ 367.58)} & 0.85\\
    \hline
    \makecell{Test set \\ (100 wells)} & \makecell{3058.42 \\ (3302.32 - 611.48 + \\ 367.58)} & 0.75 & \makecell{439.43 \\ (259.16 - 187.31 + \\ 367.58)}  & 0.82\\
    \hline
    \makecell{Robust test set \\ (90 wells)} & \makecell{2024.80 \\ (2245.71 - 551.73 + \\ 330.82)} & 0.64 & \makecell{345.16 \\ (189.99 - 175.65 + \\ 330.82)} & 0.68\\
\hline
\end{tabular}
\end{center}
\caption{Training and cross-validation statistics (defined in section \ref{crossvalidation}) computed on different normalized evaluation sets.}
\label{table:crossValidationMetric}
\end{table}

\begin{figure}[!ht]
    \centering
    \includegraphics[scale=0.6]{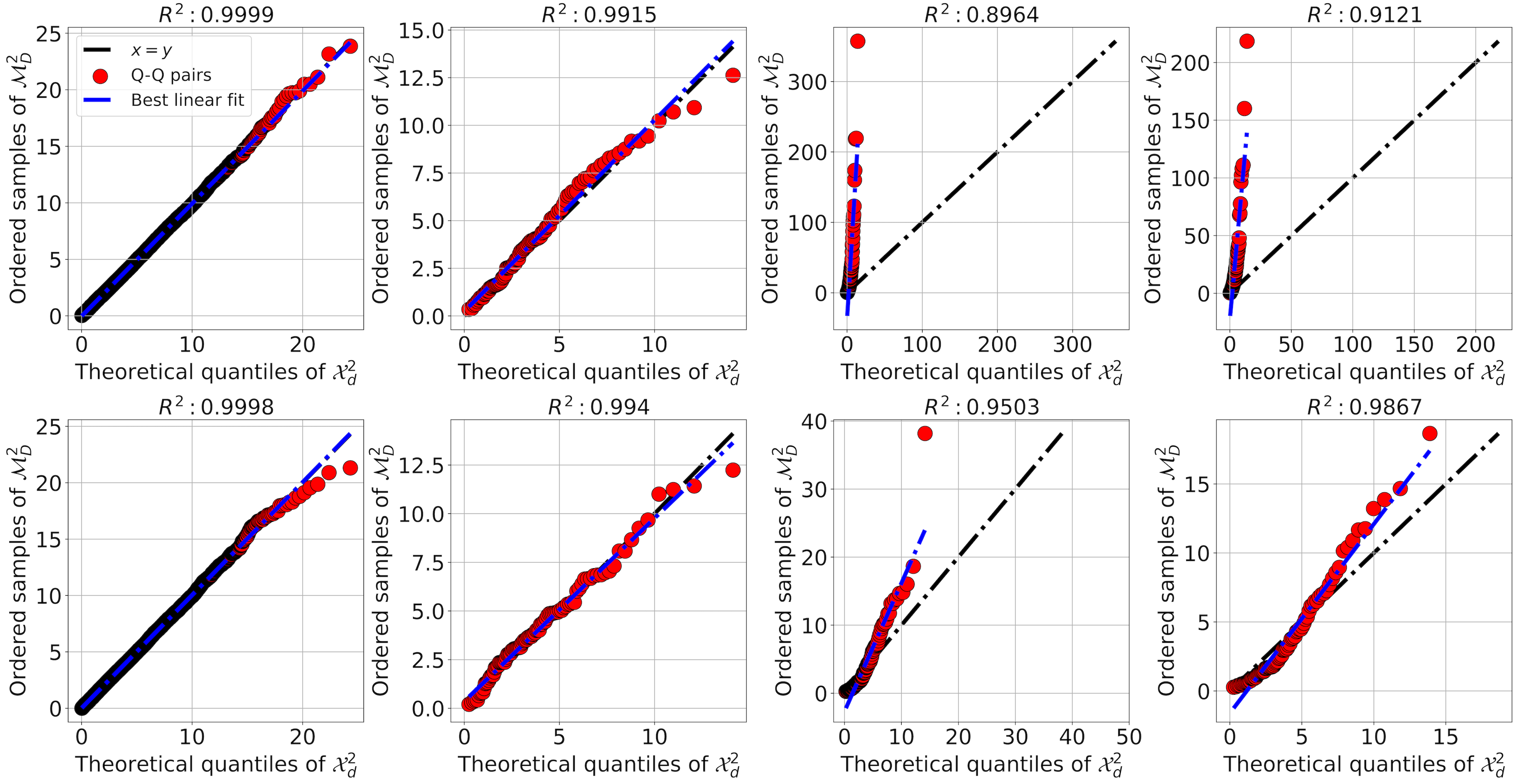}
    \caption{Chi-square Q-Q plots for testing GP (top row) vs. GP-DNN (bottom row) predictive posteriors. Empirical quantiles in the left and second from left columns are derived using 100 and 1 random samples respectively from predictive distribution at each well location in the test set. Empirical quantiles in the second from right and right columns are derived using real data in the test and robust test sets respectively. The $R^2$ coefficient of determination of the best linear fit (blue line) is listed on top.}
    \label{fig:qqplots1}
\end{figure}

In Figure \ref{fig:qqplots1}, we compare chi-square Q-Q plots to determine the fidelity of the GP and GP-DNN uncertainty estimates towards explaining the uncertainty exhibited in the blind well test set $\mathcal{T} = \{{\bm{x}_*}_1, {\bm{x}_*}_2, \ldots, {\bm{x}_*}_{m_*}\}$. Consider the hypothetical scenario in which observations $\bm{y}({\bm{x}_*}_i) \in \mathbb{R}^{d=4}, \forall {\bm{x}_*}_i \in \mathcal{T},$ are true samples from the predictive posterior distribution $a({\bm{x}_*}_i)|X_\tau,\bm{y}_\tau,{\bm{x}_*}_i$ (equation \ref{eq:cokriging}). The Q-Q plot would then result in an identity relation as discussed in section \ref{crossvalidation}. We verify this claim in the left column of Figure \ref{fig:qqplots1} where the empirical quantiles on the y-axis are estimated using 100 random samples from $a({\bm{x}_*}_i)|X_\tau,\bm{y}_\tau,{\bm{x}_*}_i, \forall {\bm{x}_*}_i \in \mathcal{T}$. The Q-Q plot shows an almost perfect identity relation. In our problem setup, only one observation per ${\bm{x}_*}_i$ is available. The effect of this limited sample size is explored in the second column from left in Figure \ref{fig:qqplots1} where one sample per $a({\bm{x}_*}_i)|X_\tau,\bm{y}_\tau,{\bm{x}_*}_i$ is utilized. The Q-Q scatter plot shows minor deviations around the identity line, with appreciable deviations generally observed for $\mathcal{M}_D^2>10$, with the value 10 corresponding to the 96\% quantile of $\mathcal{X}_{d=4}^2$ distribution. As highlighted by \cite{johnson07}, such Q-Q deviations at tail ends of the distributions become exacerbated due to limited sample size. 

In the second column from right of Figure \ref{fig:qqplots1}, the empirical quantiles on y-axis are derived using the real observations $\{\bm{y}({\bm{x}_*}_i); \forall {\bm{x}_*}_i \in \mathcal{T}\}$ available at the blind test set wells. The GP-DNN Q-Q scatter points roughly plot along the identity line for $\mathcal{M}_D^2<8$, with the value 8 corresponding to the 90\% quantile of $\mathcal{X}_{d=4}^2$ distribution. Few clear outliers with respect to the predictive posterior are also observed. The deviation from the identity relation is severe for the baseline GP Q-Q plot. Note that many empirical samples of $\mathcal{M}_D^2$ (33 out of 100 wells) even fall above the 99.99\% quantile of $\mathcal{X}_{d=4}^2$ distribution which corresponds a value of 23.5. This highlights that it is extremely unlikely that these samples belong to the baseline GP predictive posterior. Two potential factors that could be contributing to the lack of a clear one-to-one correspondence for the baseline GP model are (1) inaccurate regression estimates for the posterior uncertainty, and (2) real data noise. With regards to the latter, certain additional wells with suspect data were identified from the Q-Q plot outliers. For instance, the well with $\mathcal{M}_D^2\approx40$ (see the bottom plot on second column from right) has about 11 data samples indicating water levels declined by 170 feet from 2015 to 2017, and regained back 170 feet from 2017 to 2019, leading to a fitted seasonal oscillation amplitude value of 81 feet. This seems a clear data outlier considering the general distribution for amplitudes observed in Figure \ref{fig:targets_well}. To minimize impacts of similar outlying data samples for the Q-Q plot analysis, we created an additional test set, termed the robust test set $\mathcal{T}_R$ as described below.

We conducted an additional outlier detection exercise by utilizing the robust sample covariance based Mahalanobis-distance outlier detection as proposed by \citet{rousseeuw99}. We underscore that this outlier detection is solely based on the observed data samples and does not include any regression model covariance estimates. Specifically, the sample mean and robust sample covariances are determined from the observations of $\bm{y}$ available at all training, validation and test set wells. Subsequently, the Mahalanobis distance of samples of $\bm{y}$ are computed using the sample mean and robust sample covariance matrix. Any test well observation $\bm{y}$ which exceeds the outlier detection threshold, set to be the 99\% quantile of $\mathcal{X}_{d=4}^2$, are flagged as outliers with respect to the sample data distribution. Following this procedure, 10 test set wells were discarded for having data outliers. In the rightmost column of Figure \ref{fig:qqplots1}, we compare the Q-Q plots of baseline GP and GP-DNN models obtained using $\mathcal{T}_R$. While the linearity of the Q-Q trend slightly improved (compare $R^2$ coefficient of determination for the fit of the data to the blue line), the baseline GP model still exhibits significant deviations from the identity relation reinforcing the claim that baseline GP model yielded inaccurate posterior uncertainty estimates. For the GP-DNN model, the $R^2$ coefficient of the best linear fit to the Q-Q trend is 98.67\% indicating that the shape of the empirical and theoretical distributions are practically identical. Approximate one-to-one correspondence between empirical and theoretical distributions are observed up to the 90\% quantile ($\approx8$ for $\mathcal{X}_{d=4}^2$). Noticeable mismatches between the empirical and theoretical quantities especially occur beyond the 90\% quantile of $\mathcal{X}_{d=4}^2$. Such deviations are expected given the limited sample size at each test location as well as due to the simplified modeling assumptions made regarding the data noise, e.g., spatially uniform noise variance. 

\subsection{Results} \label{Results}
In this section, we will discuss estimated water level trends and associated uncertainties in the CV during 2015-2020, as inferred by two GP regression models. We show that the mean field estimates from both models show somewhat comparable spatial variability. However, the results interpretation will be primarily focused on the GP-DNN results since corresponding posterior predictive distribution is able to explain the held out test set with higher log-likelihood statistics and robust predictive posterior uncertainty estimates as discussed in the previous section. The GP-DNN model is able to model the uncertainty more reliably because of its ability to handle non-stationary data in the latent space as shown in section \ref{latent space}.

\subsubsection{Mean groundwater levels during 2015-2020} \label{result_mean}
\begin{figure}[!ht]
    \centering
    \includegraphics[scale=1.5]{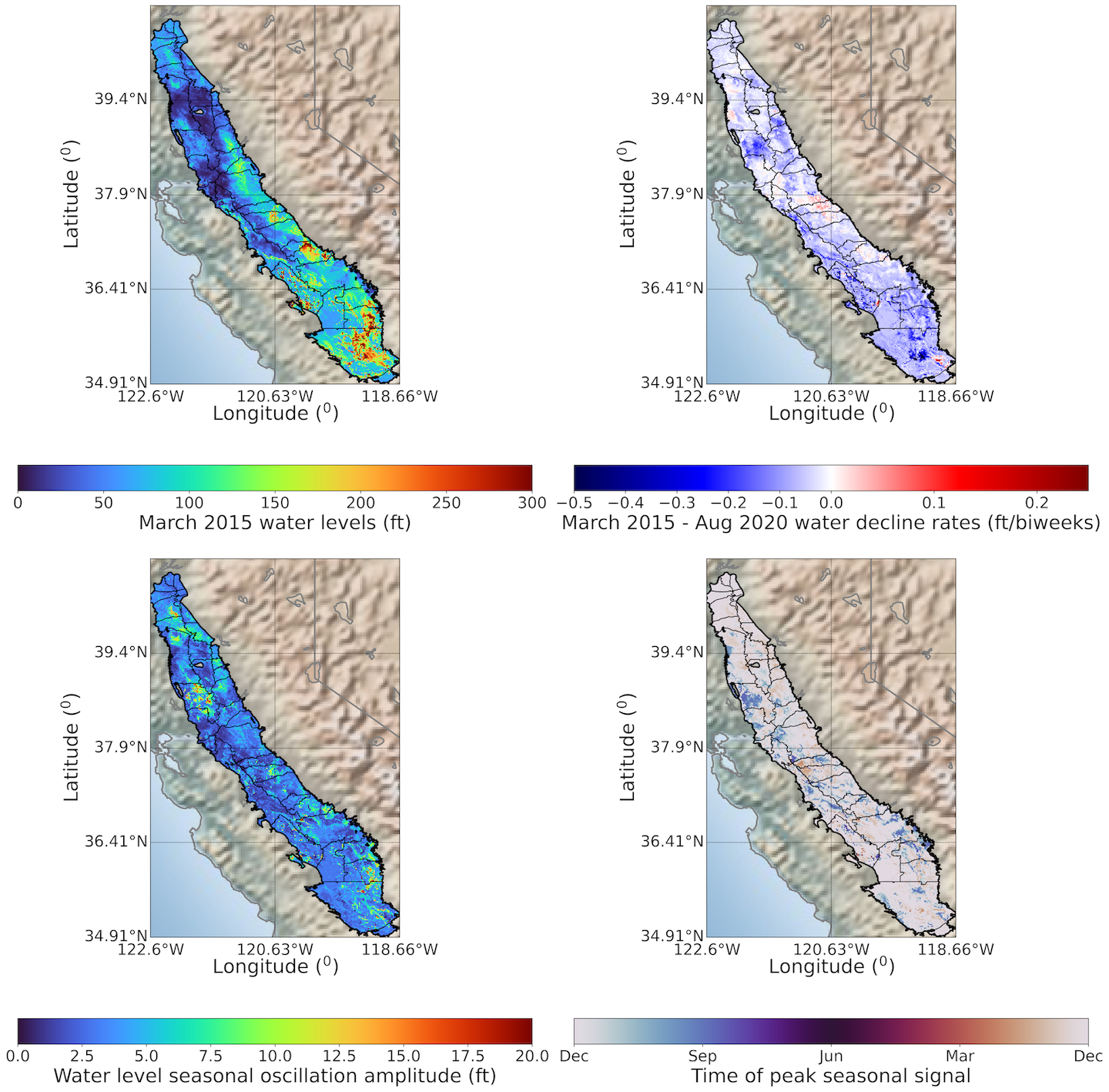}
    \caption{Mean of the posterior predictive distribution of water level long-term and seasonal trend parameters predicted by GP-DNN regression. The plots, clockwise from top left, correspond to $a_1(\bm{x})$, $a_2(\bm{x})$, $a_3(\bm{x})$ and $a_4(\bm{x})$. For $a_2(\bm{x})$, blue indicates rising water levels.}
    \label{fig:posterior_mean_dnn}
\end{figure}

Mean water level trends predicted by the GP-DNN model are shown in Figure \ref{fig:posterior_mean_dnn}. In March 2015, water levels were predicted to be significantly shallower (upto 50 feet from surface) in the SV. The SJV, on the other hands shows greater variability. Along the north-western flank of the SJV (East Conta Costa, Tracy, Delta Mendota, and western edges of Eastern San Joaquin, Modesto, Turlock and Merced subbasins), water levels within 50 feet from surface are observed. Water levels deeper than 100 feet from surface are predicted along the eastern boundary and southern half of the SJV. The deepest water levels ($>$ 200 feet) are observed across several isolated regions in the Modesto, Turlock, Chowchilla, Madera, Westside, Kaweah, Tule and Kern counties. The spatial continuity of the patterns manifested in well data (Figure \ref{fig:targets_well}) has been preserved in the estimated mean fields. Larger-scale spatially correlated structures are observed in the central SV and northern SJV, while rapidly varying spatial patterns (short length-scales) are observed in southern SJV area. In a previous study by \cite{gualandi21}, shallow aquifer (aquifer layers above the clay confining unit) processes in the SJV were found to be contributing factors to these short length-scale variations in aquifer responses.

\begin{figure}[!ht]
    \centering
    \includegraphics[scale=1.1]{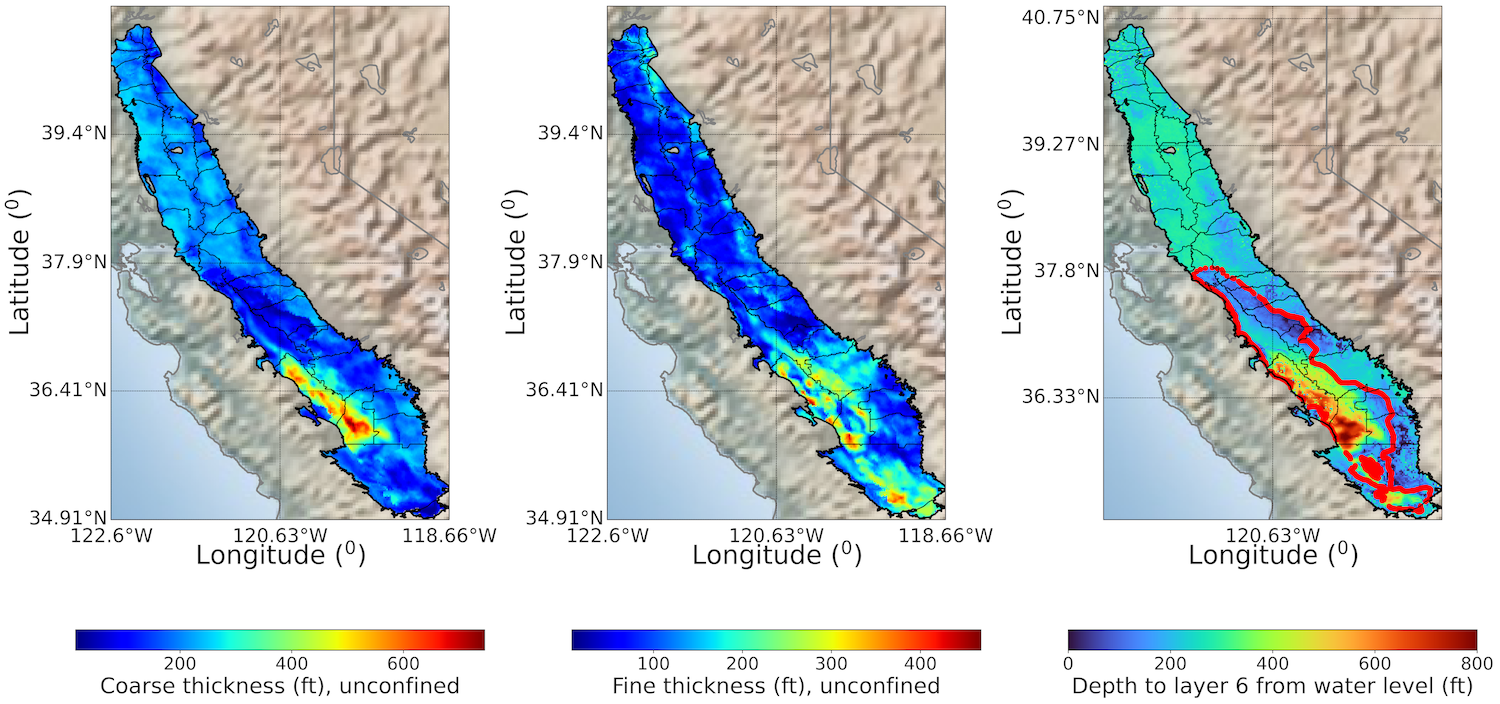}
    \caption{Total thickness of coarse (left) and fine grained sediments (middle) in the upper semi-confined aquifer (texture model layers 1-5). (Right) Depth thickness between predicted March 2015 water level mean and the top the $6^{th}$ layer. Extent of Corcoran clay is shown in red.}
    \label{fig:total_coarse_fine}
\end{figure}

During 2015-2020, water levels have exhibited both positive and negative decline trends, with a large proportion of the locations showing sustainable changes in groundwater levels. On average, 98\% of locations in the CV grid have had moderate fluctuations in the groundwater levels, ranging between $\pm$ 12 feet during 2015-2020 (decline rates ranging between $\pm$ 0.1 ft/biweeks). 11\% of the locations have witnessed uplifts exceeding 12 feet, while 1\% of the locations underwent declines exceeding 12 feet. While CV groundwater reservoirs continually experienced groundwater loss during the 2012-2015 drought (\cite{ojha19}, \cite{liu22}), our results indicate that, on average, there were few locations with large declines in groundwater ($>25$ feet) during 2015-2020. This is likely due to the exceptionally wet years of 2017 and 2019 that have resulted in partial, localized recovery of water levels as reported in several studies (\cite{dwr17}, \cite{dwr19}). However, note that these short recharge periods have been interspersed with prolonged periods of drought, in what has been termed as a megadrought, and current CV groundwater levels in general lie significantly below the pre-2006 drought levels (\cite{liu22}). Approximately 63\% of the 11\% CV locations that witnessed appreciable uplifts in water levels include locations in the Tulare Lake hydrological basin (TLHB; includes Westside, Kings, Kaweah, Tulare Lake, Tule and Kern county subbasins). These results align with the \cite{dwr19} report that shows several wells in the TLHB observed uplifts during Spring 2016 - Spring 2019, with 31\% of the 624 wells logged ranging between 5 to 25 feet uplifts and 25\% exceeding 25 feet. \cite{neely2021} also presented similar observations while studying groundwater depletion related surface deformation with remote sensing data in CV. They observed strong surface uplift in Westside during the 2017 wet year potentially due to the above average aquifer recharge in that year. We hypothesize two possible reasons for these observed uplifts in water level mean fields.
\begin{enumerate}
    \item \emph{Underlying hydrogeology}: Using the sediment thickness features presented in section \ref{features}, we observed that the western flank of the TLHB contains some of the thickest coarse-grained sediment columns in the semi-confined aquifer zone. Shown in Figure \ref{fig:total_coarse_fine} are the total thickness of coarse and fine-grained sediments in the upper semi-confined aquifer (layers 1-5 of the CV texture model). It may be observed that the thickness of the semi-confined coarse grained sediments in western TLBH ranges within $\approx$ 400-700 feet, significantly larger than any other region of the CV. Thick fine-grained sediments were also estimated to exist in western and southern TLHB. Also plotted is the depth to bottom of layer 6  of the texture model, corresponding to the top of the Corcoran Clay where it exists, from the GP-DNN predicted mean water level. Comparing with Figure \ref{fig:posterior_mean_dnn}, mean water levels in western TLHB are predicted to stay mostly between 50-400 feet below the surface and 200-800 feet above the Corcoran Clay. Note that in comparison the mean water levels in the SV and northern SJV exist at shallower depths ($\approx$ 0-100 feet). Given that (1) the western TLHB semi-confined aquifer is a structural trough with thick coarse grained sediments and (2) coarse-grained sediments have higher storage and hydraulic conductivity, it is possible that the higher influx of water during wet years 2017 and 2019 resulted in preferential recharge of the western TLHB aquifers. This preferential recharge effect in western TLHB has also been confirmed in independent component analyses of InSAR time series data conducted by \cite{gualandi21}. As mentioned beforehand, modeling limitation of handling groundwater flow in 2D would also likely introduce dimensionality issues, such as not considering vertical soil infiltration. Further, other factors such as varying groundwater pumping rates, delayed pressure dissipation across the Corcoran clay, structural barriers or pathways inside the stratigraphies considered in the CV texture model and the spatial heterogeneity were also not considered in this study. Many of these variables are unknown or known with large uncertainty in the CV, impeding building a robust physical 3D model for groundwater flow as discussed earlier. 
    \item \emph{Data sparsity and noise}: Well data from the TLHB were particularly sparse as compared to the rest of the study area (Figure \ref{fig:data_density}), with nearly all of the wells having $<$ 15 samples during 2015-2020. Given that we assumed a single linear model for the long-term signal over the five year period, data aliasing effects could have potentially led to noisy estimates of the water level trends at wells in the TLHB.
\end{enumerate}
Given the large uncertainty associated with the hydrogeology and well-data, it is desirable to quantify prediction uncertainty. In the next section, we discuss how the GP approach allows deriving the full posterior predictive distribution and the ability to simulate equiprobable estimates of the water level trends. 

Regarding the mean seasonal signal, 95\% of the locations are predicted to have small ($<$ 10 feet) seasonal peak-to-peak oscillations. In the SV, most of the larger oscillations are observed in Red Bluff, Corning, Vina, North Yuba, South Yuba and Yolo subbasins. In the SJV, larger amplitude seasonal variability occur regionally in the Modesto, Turlock, Chowchilla, Madera, Tule and Kern counties. Based on the phase delay field, timing of the peak seasonal oscillations are predicted to occur between October to April at 93\% of the CV locations. The peak seasonal signal is controlled by various factors such as groundwater production, precipitation, snow runoff and ease of surface water infiltration into the aquifer (\cite{riel2018}). The predicted mean October to April peak generally correlates with wet months of the seasonal cycle and thus decreased groundwater production in the valley.  Note that accurately estimating the timing of the peak seasonal signal will be challenging given the well data are very sparsely sampled. The GP-DNN posterior estimates appropriately capture this modeling uncertainty as discussed in sections \ref{training} and \ref{nonstationarity_uncertainty}.

\begin{figure}[!ht]
    \centering
    \includegraphics[scale=1.5]{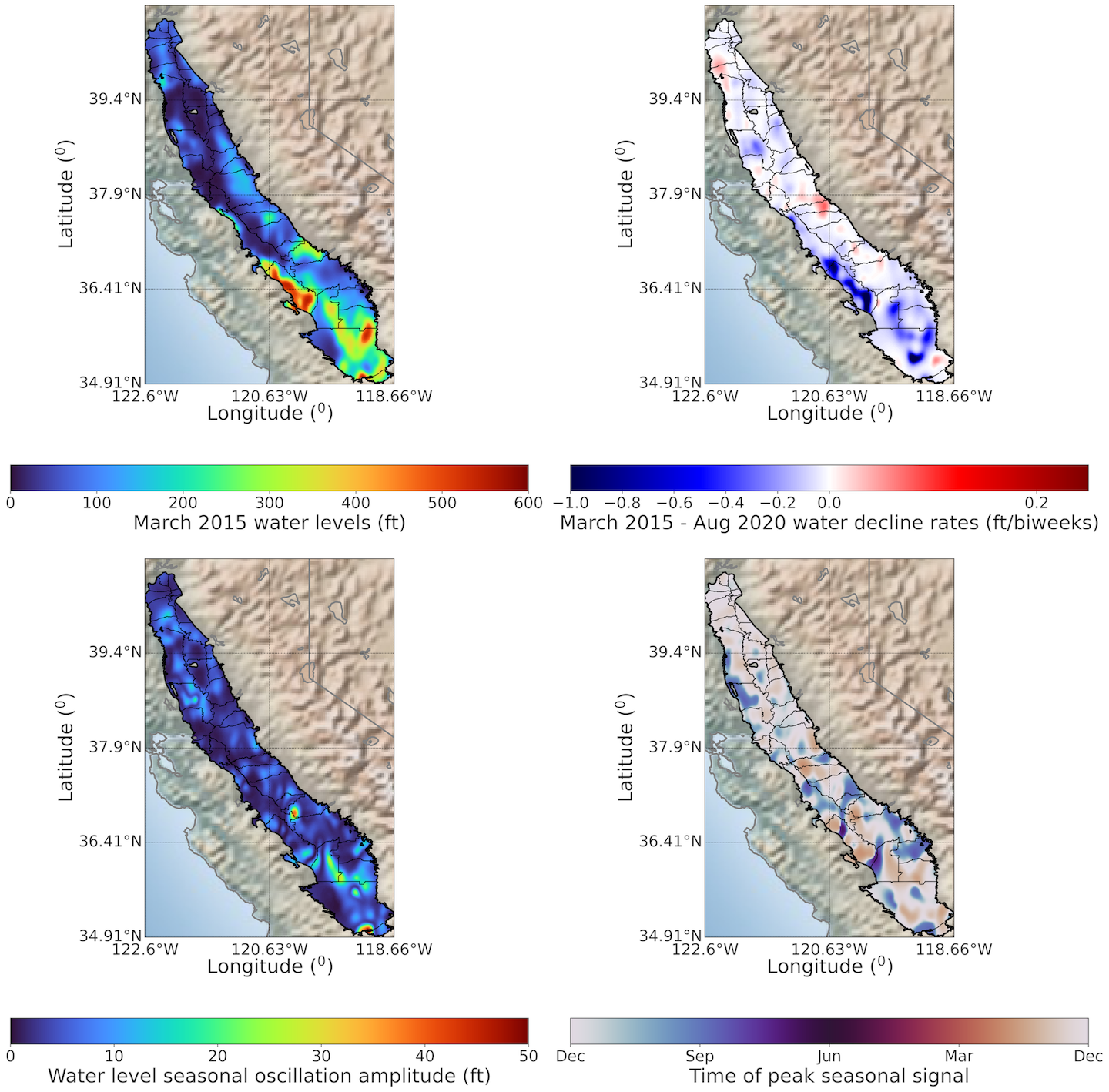}
    \caption{Mean of the posterior predictive distribution of water level long-term and seasonal trend parameters predicted by baseline GP regression.}
    \label{fig:posterior_mean_cokrig}
\end{figure}

Figure \ref{fig:posterior_mean_cokrig} shows the mean fields predicted by the baseline GP regression approach. While the fields generally show coarse correlation with the GP-DNN mean (Figure \ref{fig:posterior_mean_dnn}), the model in general tends to predict stationary correlations in the study area. This is immediately apparent in the March 2015 water levels in Westside and Tulare Lake subbasins. The rough variability manifested in the well data (Figure \ref{fig:targets_well}) have been smoothed given a stationary kernel is used to smooth the well data. Similar high amplitude artefacts in southern SJV are also observed in the seasonal amplitude map. The GP-DNN model, on the other hand, does not suffer from this stationary smoothing limitation and is able to capture both large-scale and fine-scale variability in the data as shown previously.

\subsubsection{Uncertainty and non-stationarity in groundwater levels} \label{nonstationarity_uncertainty}

\begin{figure}[!ht]
    \centering
    \includegraphics[scale=1.75]{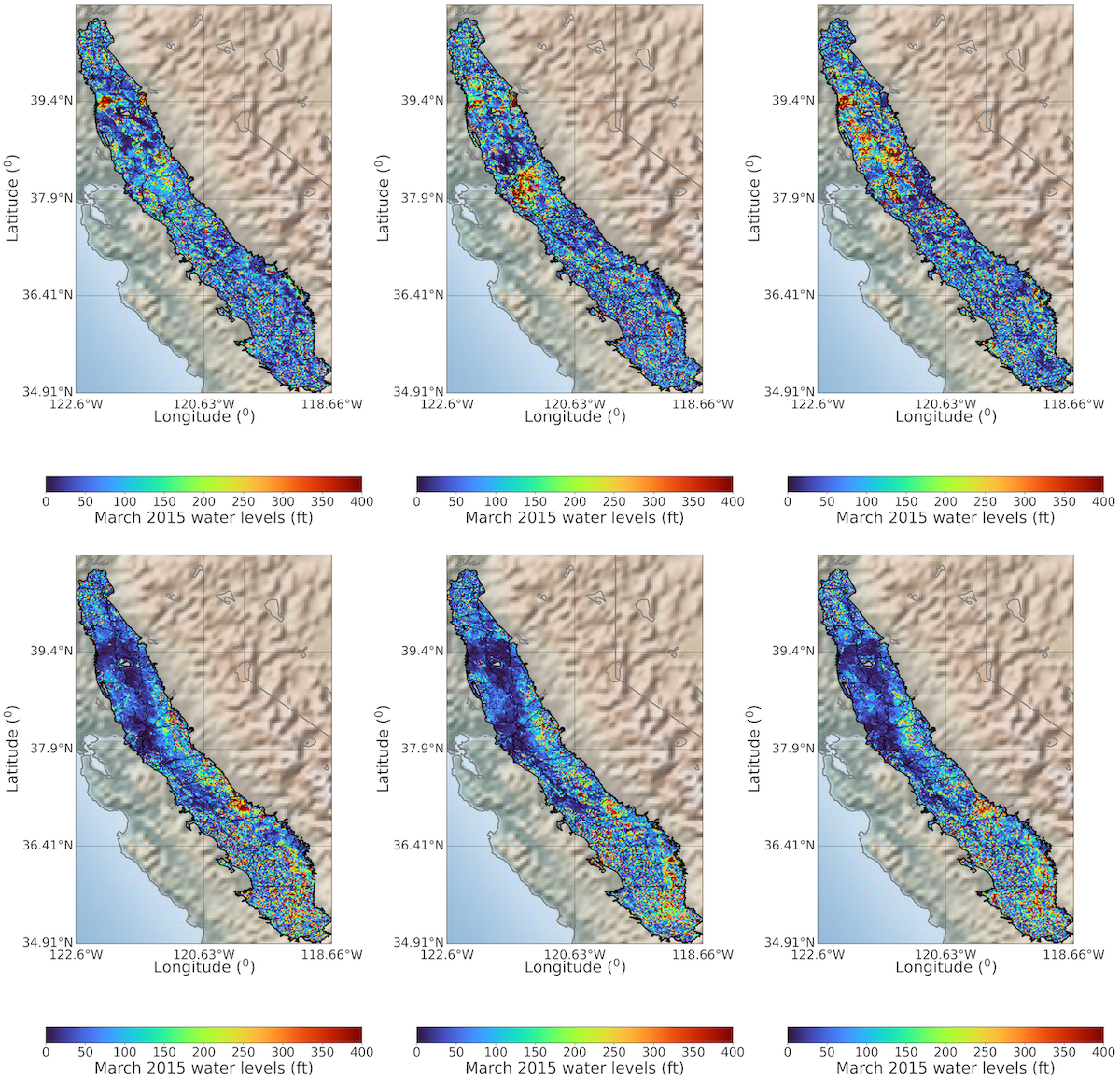}
    \caption{Three random realizations of water levels in March 2015, sampled from the GP-DNN prior (top row) and posterior (bottom row) predictive distributions.}
    \label{fig:prior_realizations_S1_S2_intercept}
\end{figure}

Random realizations of March 2015 water levels from the GP-DNN distribution prior to well data conditioning  (Equation \ref{eq:DNNGPPrior}) are shown in Figure \ref{fig:prior_realizations_S1_S2_intercept}. Non-stationary spatial heterogeneity of the simulated patterns across the valley is immediately apparent. Starting from central SV (Colusa, Butte and Vina subbasins) till the northern border of the SJV (Tracy and Eastern San Joaquin subbasins), large length-scale structures were simulated. Variability of water levels occurs along medium range structures in regions without the Corcoran clay confining unit such as the eastern flanks of the Modesto, Turlock, Merced, Madera, Kings, Kaweah, Tule and Kern County subbasins. On the other hand, regions located above the confining unit exhibit very fine length-scale variations, likely related to shallow semi-confined aquifer responses. The prior predictive uncertainty was conditioned to training data, yielding the posterior predictive distribution (equation \ref{eq:cokriging}). Posterior predictive realizations are also shown in Figure \ref{fig:prior_realizations_S1_S2_intercept}. Similar to the prior realizations, the GP-DNN posterior predictive samples exhibit non-stationary spatial patterns across the modeling domain. The self-consistency of the three posterior predictive samples stands in contrast with the more highly-variable samples drawn from the prior. This indicates that conditioning is, in many places within the domain, effectively constraining the prior when equation \ref{eq:cokriging} is applied. 

\begin{figure}[!ht]
    \centering
    \includegraphics[scale=0.37]{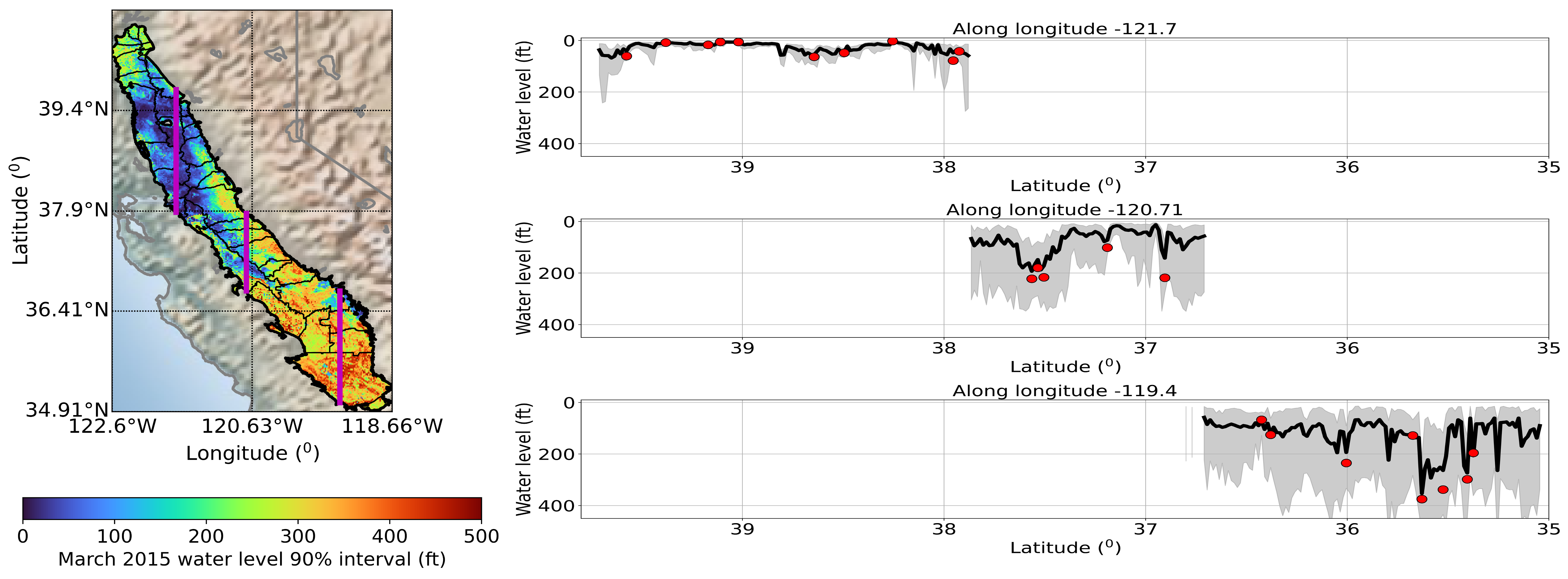}
    \caption{(Left) P10-P90 uncertainty interval of March 2015 water level computed using 500 posterior predictive samples from GP-DNN model. (Right) GP-DNN predictions along three different longitude transects highlighted in magenta on the left plot. Black line: posterior predictive mean; gray: P10-P90; red circles: well data.}
    \label{fig:1dprofiles_GPDNN}
\end{figure}

\begin{figure}[!ht]
    \centering
    \includegraphics[scale=0.37]{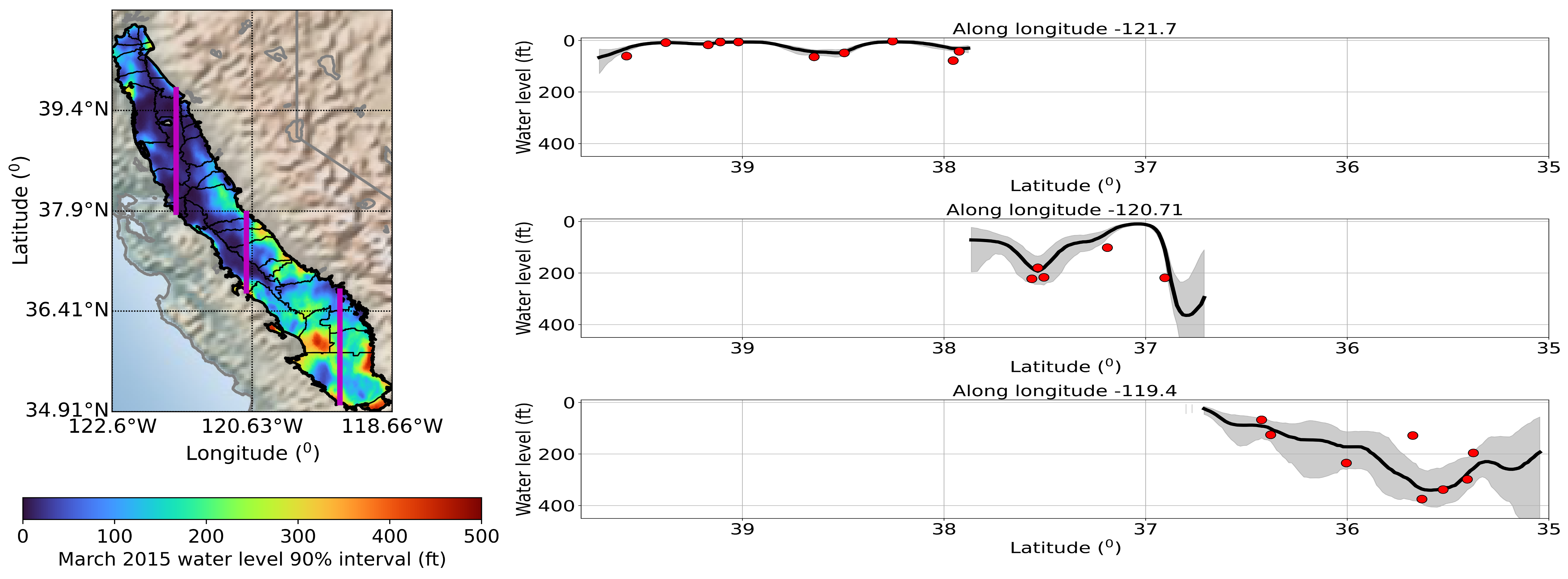}
    \caption{(Left) P10-P90 uncertainty interval of March 2015 water level computed using 500 posterior predictive samples from baseline GP model. (Right) Baseline GP predictions along three different longitude transects highlighted in magenta on the left plot. Black line: posterior predictive mean; gray: P10-P90; red circles: well data.}
    \label{fig:1dprofiles_GP}
\end{figure}

\begin{figure}[!ht]
    \centering
    \includegraphics[scale=0.28]{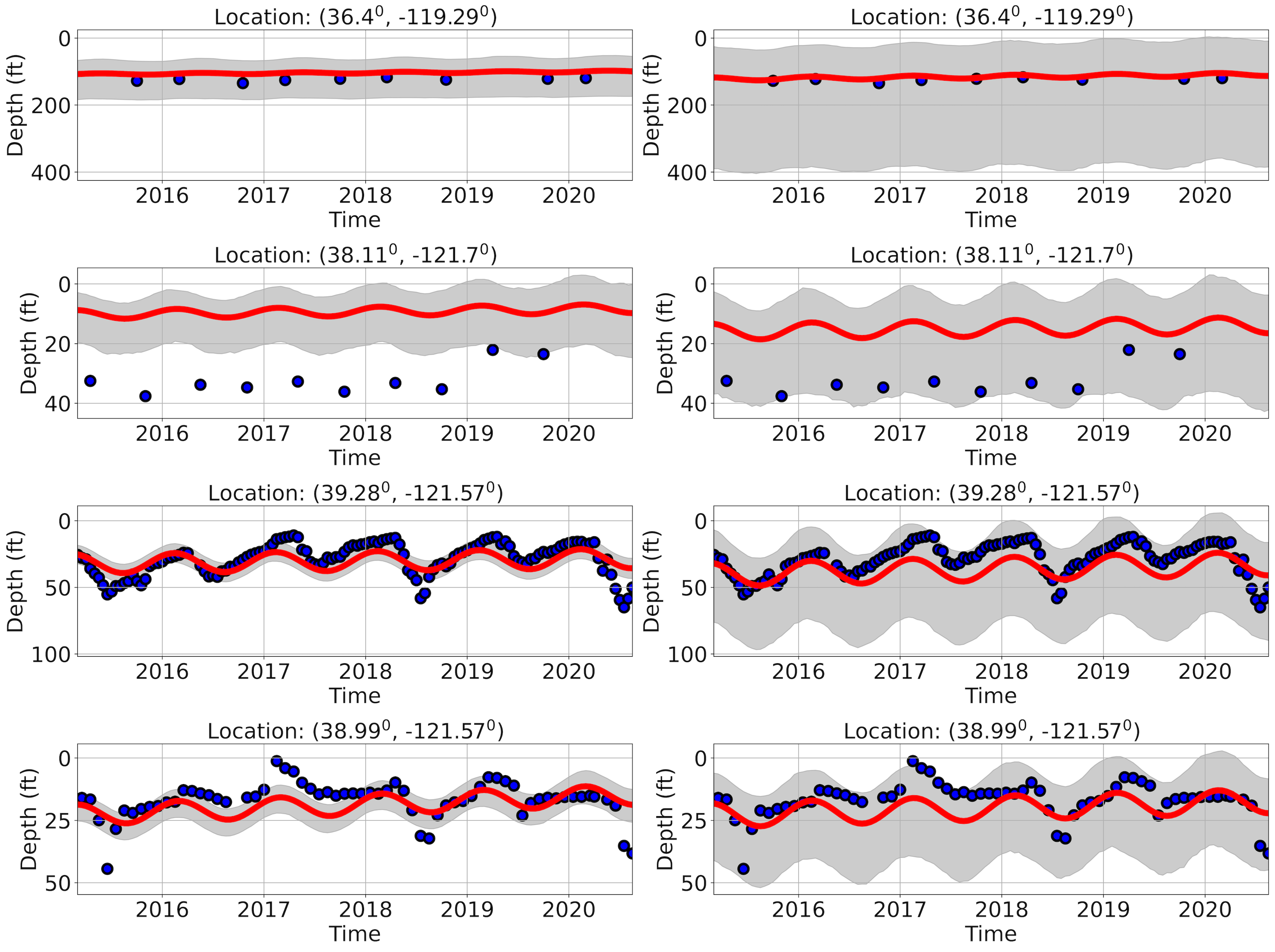}
    \caption{Modeled mean water level depths (red line), P10-P90 uncertainty intervals computed using 500 posterior predictive samples and observed data at four blind wells predicted using baseline GP (left column) and GP-DNN (right column) models.}
    \label{fig:blind_well_wl}
\end{figure}

Figure \ref{fig:1dprofiles_GPDNN} shows the sample standard deviation associated with March 2015 water level mean, computed using 500 random samples from the GP-DNN posterior predictive distribution. The corresponding results for basline GP case are shown in Figure \ref{fig:1dprofiles_GP} for comparison. Result interpretation is mostly focused on GP-DNN results as it was shown previously that basline GP model predictions are inaccurate (see cross-validation in section \ref{training}).  We observe tight uncertainty intervals in central to southern SV and in most areas in the north-western SJV. Wider uncertainty intervals are primarily predicted for locations overlying the clay confining unit and along the northern and eastern domain boundaries. Note that the posterior predictive covariance (equation \ref{eq:cokriging}) depends only on the training covariance matrix $K_{\tau\tau}$, testing covariance matrix $K_{**}$ and the training-testing covariance matrix $K_{\tau*}$. From equation \ref{eq:stationaryKernel}, it follows that the GP-DNN uncertainty predictions are in accordance with distances from training and testing locations as mapped in the latent space. Locations above the Corcoran clay have significant differences in the semi-confined aquifer structural and lithological properties, thus leading to greater separation in the latent space (see section \ref{latent space}) and consequently larger uncertainty. We underscore that the well data itself shows very rapid and noisy variability in this area (which could be due to hydrogeological heterogenity, observational noise and other factors as discussed in section \ref{result_mean}). As described previously, we have plotted the posterior predictive mean and 500 samples along three longitude transects and overlain the well data. Visually, well data from the northern CV show wider correlated patterns as compared to the south where the correlation falls off very rapidly with distance. This is especially true of the bottom transect passing through Westside subbasin where we see fluctuations ranging within $\pm 300$ feet roughly across a distance of 60 miles. The GP-DNN model is able to replicate this abrupt data variability within the modeling domain by latent space reconfiguration and does not force the simulations to be overly smooth, which is a well-documented limitation associated with kriging type models for modeling geological heterogeneity (\cite{linde2015}). Figure \ref{fig:1dprofiles_GP} shows the effect of smoothed mean estimates for the baseline GP model. Also, note that predicted uncertainty is driven by conditioning data proximity as expected. However, a limitation with the GP-DNN regression is that some of the posterior predictive samples demonstrate very noisy variability in the regions above the Corcoran clay. In section \ref{discussion}, we discuss this issue in greater detail and propose ideas for future work. The linear model shown in equation \ref{eq:linearReg1} may be used to obtain a spatio-temporally continuous mean water level and associated variance. Figure \ref{fig:blind_well_wl} compares the modeled mean water level and P10-P90 uncertainty intervals computed at four selected blind wells for the two regression models under study. While the mean water level predictions are more or less similar, the uncertainty intervals may vary significantly across the two wells. The GP modeling may lead to overly confident uncertainty estimates, leading to the real data lying outside the predicted P10-P90 intervals (see second well from top in Figure \ref{fig:blind_well_wl}). By predicting wider uncertainty intervals in some cases, the GP-DNN model is able to appropriately capture the uncertainty existing due to data unavabaility and noise.

\subsubsection{GP-DNN latent space interpretation} \label{latent space}
While deep learning models provide the flexibility to approximate very complex non-linear relations, physically explaining the model predictions, for instance understanding how the input features get propagated through the model into outputs, has remained a challenge and is an active area of research (\cite{samek21}). In this paper, deep neural networks were used to transform the hydrogeological features into latent features. To explain the latent space features, we perform dimension reduction of the latent space and present visual analyses of the variability of aquifer and aquitard sediment thicknesses in the latent space as described in this section. 

\begin{figure}[!ht]
    \centering
    \includegraphics[scale=1.]{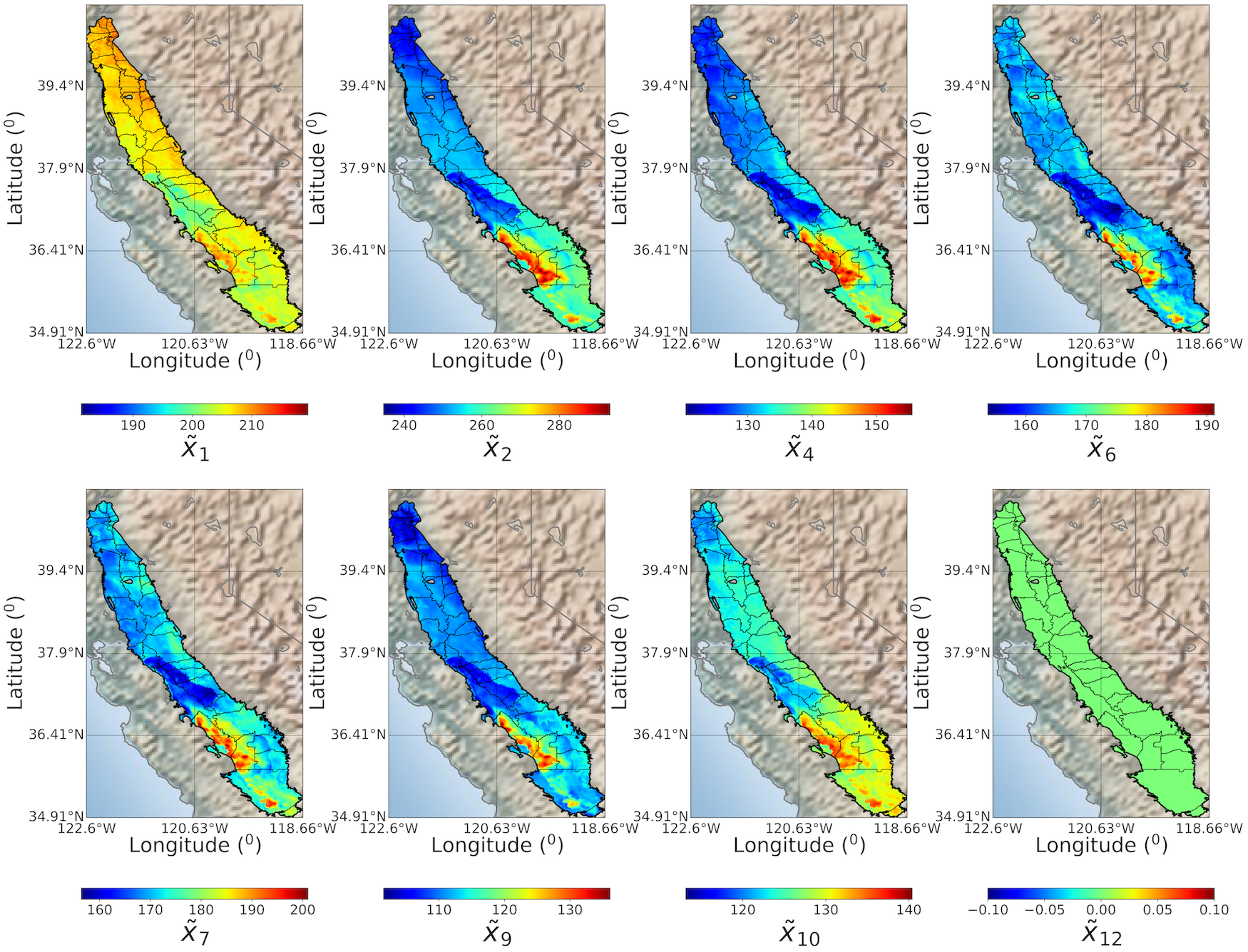}
    \caption{Selected components of the DNN output latent space $\Tilde{\bm{x}}$.}
    \label{fig:x_tilde}
\end{figure}

The dimensionality of the latent space, controlled by the number of output nodes of the DNN, was treated as a hyper-parameter and tuned to 12 by cross-validation. In Figure \ref{fig:x_tilde}, we show selected components of $\Tilde{\bm{x}}$ in map format, i.e., plotting them versus $\bm{x}$. While it is not immediately straight-forward to quantitatively interpret the latent fields from a hydrogeological perspective, we put forth some qualitative observations. The footprint of the Corcoran clay unit can be observed along the majority of the latent dimensions. Many of the displayed latent fields show similar high amplitude patterns in the Westside, Tulare Lake and Kern County subbasins. We found that these patterns exhibit a high correlation with the coarse and fine grained sediment thicknesses in the upper semi-confined aquifer. Similar high amplitude patterns may be observed for the sediment thicknesses in the upper semi-confined aquifer (Figure \ref{fig:total_coarse_fine}). High amplitude structures in $\Tilde{x}_{6}$ and $\Tilde{x}_{7}$ (for instance in North and South Yuba subbasins) correspond with areas with thick clay columns in the lower semi-confined aquifer zone. These observations generally indicate that the latent space reconfiguration was optimized to differentiate between locations based on their underlying lithological texture. 

\begin{figure}[!ht]
    \centering
    \includegraphics[scale=0.85]{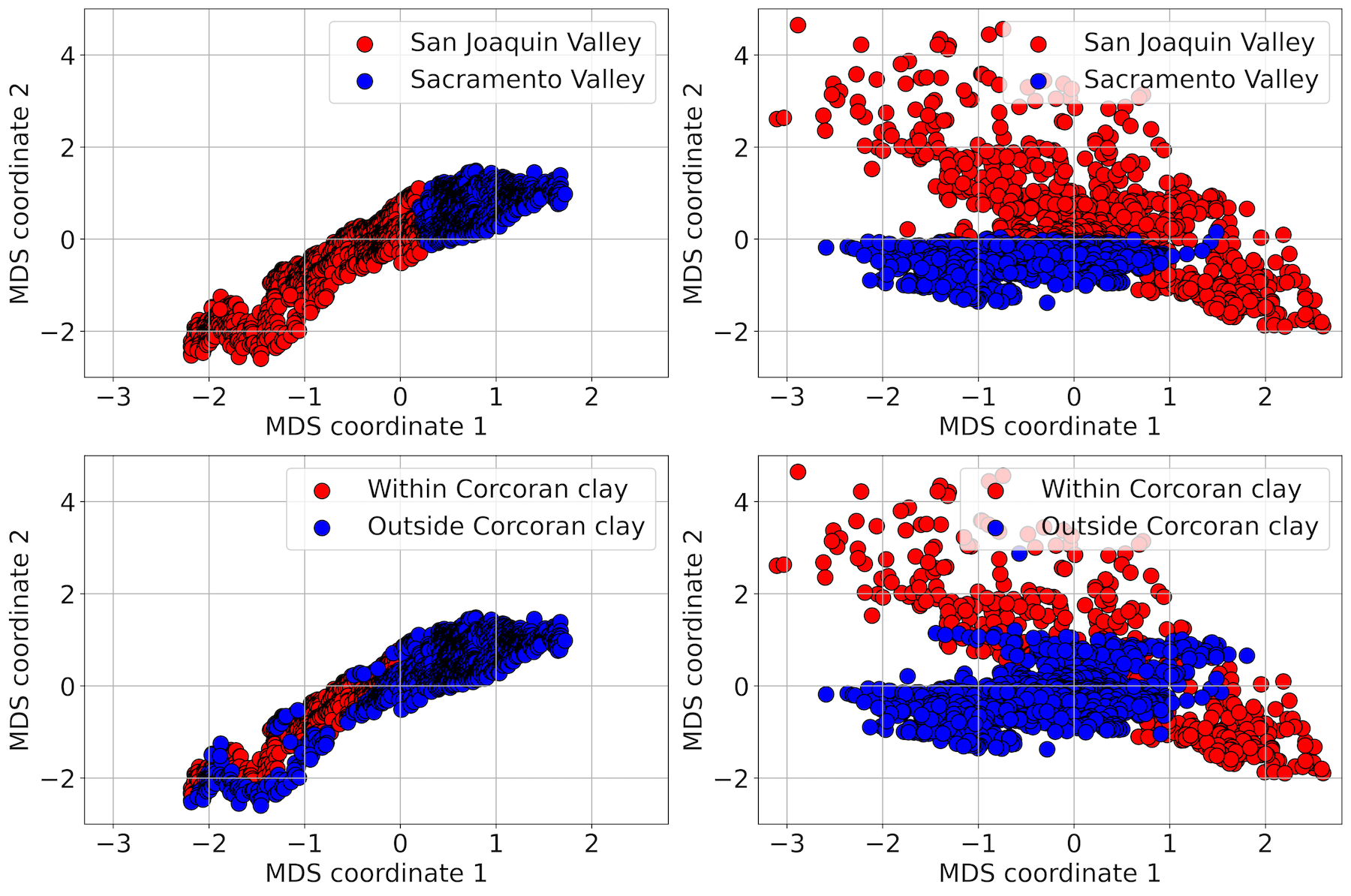}
    \caption{Training samples in compressed dimensions after MDS from $\bm{x}$-space (left) and $\bm{\Tilde{\bm{x}}}$-space (right).} 
    \label{fig:mds_valleys_clay}
\end{figure}

It is impossible to fully visualize and interpret how the latent space re-configuration varies with the texture properties in a 12-dimensional output space. Hence, we employ dimension reduction of the latent feature space to facilitate visual analysis. Specifically, we employ multidimensional scaling (MDS; \cite{borg97}), which projects samples from an input feature space into a lower-dimensional space while ensuring that the mutual distances between samples in the original space are preserved. This property of MDS is especially desirable since the covariance kernel is specified as a function of $\|\Tilde{\bm{x}}-\Tilde{\bm{x}}'\|_2$. MDS has been employed for visualization of spatial subsurface uncertainty in several previous studies (\cite{scheidt09}, \cite{pradhan20b}, \cite{pradhan22}). We compress the sample locations from the $\Tilde{\bm{x}}$-space into a two dimensional space by metric MDS. The distance measure between samples $\Tilde{\bm{x}}$ and $\Tilde{\bm{x}}'$ is taken to be $\|\Tilde{\bm{x}}-\Tilde{\bm{x}}'\|_2$ which will be preserved during dimension reduction. Figure \ref{fig:mds_valleys_clay} compares the mutual distances between the training well locations in the $\bm{x}$-space against the $\Tilde{\bm{x}}$-space after MDS. To support visual analysis, the $\Tilde{\bm{x}}$-MDS and $\bm{x}$-MDS coordinates were separately normalized to have zero mean and unit variance. Note that the relative configuration of the samples along the MDS coordinates is representative with high accuracy of the relative configuration in the latent space. The correlation coefficient between Euclidean distances in the input feature space and Euclidean distances in the MDS space was calculated to be nearly 98\%, indicating minimal loss of information during MDS. In subsequent discussion, references to the input feature space and its corresponding MDS space are used interchangeably.

\begin{figure}[!ht]
    \centering
    \includegraphics[scale=0.85]{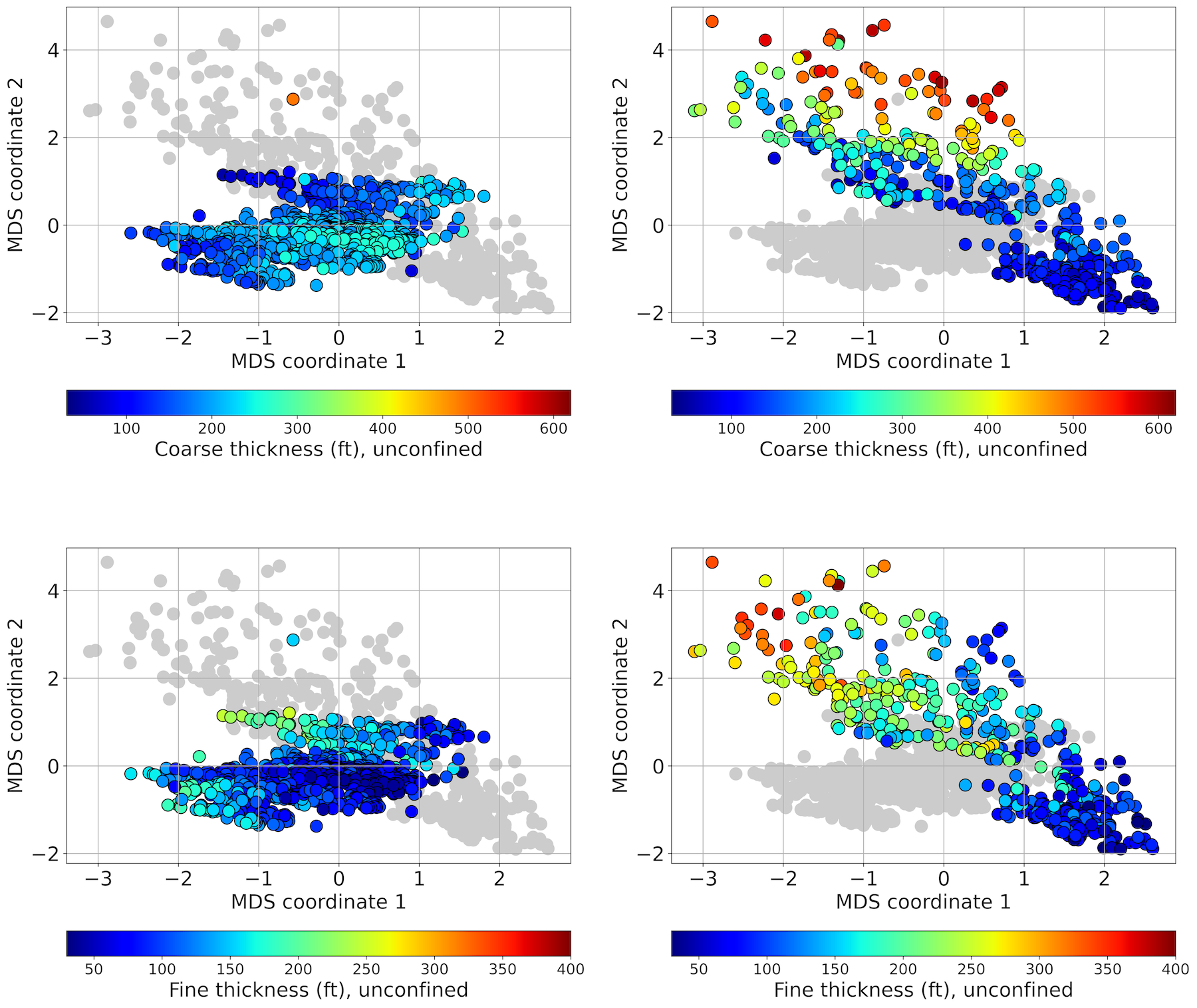}
    \caption{Training wells in $\Tilde{\bm{x}}$-MDS space colored by thickness of coarse-grained sediments (top row) and fine-grained sediments (bottom row) in the shallow semi-confined aquifer zone. Locations with and without the Corcoran clay have been grayed out in the left and right columns respectively.} 
    \label{fig:mds_tilde_norm_coarse_conf}
\end{figure}

In Figure \ref{fig:mds_valleys_clay}, the samples have been colored by their native hydrological valley and whether the Corocoran Clay exists in the subsurface. It is immediately apparent that the DNN has learned to nicely separate based on the underlying aquifer textures. In the geospatial domain, locations without the clay confining unit exist both in the SV and eastern SJV. In the latent space, these sites have been gathered into a tight cluster. This explains the corresponding tighter confidence intervals observed in Figure \ref{fig:1dprofiles_GPDNN} (compare top longitude transect against the bottom). Locations overlying the confining clay unit have been reconfigured with a distinct quasi-linear trend with a northwest to southeast orientation, with the semi-confined fine grained sediment equivalent thickness exhibiting a smooth gradation along this trend. The corresponding coarse grained sediment thicknesses also exhibit a clear southwest-northeast trending gradation (Figure \ref{fig:mds_tilde_norm_coarse_conf}), correlated with the scatter about the trend. Note especially that the variance about the northwest to southeast quasi-linear trend increases with in correlation with fine grained (clayey) sediment thickness.  This shows that the DNN has learnt to differentiate between locations based on their underlying equivalent coarse and fine grained sediment column heights. From first principles, this is a natural expectation since two 1D aquifer columns with equal height, homogeneous sediment type and properties, and same boundary conditions will result in equivalent pressure head profiles (equation \ref{eq:diffusion3D}). Greater the proportion of clay sediments in the aquifer column, the more heterogeneous the head profiles are expected to be. Thus, the flexibility of the GP-DNN formulation to distend and squash spatial distances facilitates it to (1) yield uncertainty predictions that are \emph{texturally-aware} instead of being just \emph{geospatially-aware}, as is typical in kriging-type models, and (2) model non-stationary spatial heterogeneity across the modeling domain. As validated with the blind well test set in section \ref{training}, this leads to reliable uncertainty quantification. 

\section{Discussion} \label{discussion}
 In this section, we discuss limitations and advantages of the GP-DNN regression model along with directions for future research. It should be noted that the CV texture model was obtained by non-stationary kriging of well data and has uncertainty associated with it. This is especially true along the sub-domain boundaries (for instance mapped edge of the Corcoran clay) and deeper aquifer where the well texture data is scant leading to artifacts in the model (\cite{faunt09}). While we did not explicitly account for this uncertainty, we found empirically that the DNN ignores noisy features and artifacts especially from the deeper sections of the texture model. To improve reliability of the model predictions, additional robust features may be considered in the future. Such features could include (1) hydrogeophysical data such as electromagnetic data (\cite{kang21}) which may provide spatially continuous information on aquifer structure and heterogeneity, (2) remote-sensing data such as InSAR surface deformation data which could inform on the sediment elastic/inelastic properties, (3) precipitation data, surface water delivery data, and crop water use data which serve as proxies for groundwater source and sink flux. With additional data, it might be necessary to consider other regression models that can handle different data structures. For example, convolutional neural networks (CNNs; \cite{krizhevsky12}) may be more suitable to learn from spatial geophysical data compared to DNNs (\cite{pradhan20c}, \cite{pradhan22}). The approach of hierarchically combining a regression model with GPs may in theory be extended to CNNs, with future effort required for applicable model design. 
 
 As discussed earlier, 3D variability of water levels resulting from vertical connectivity of aquifer layers was not considered and is a limitation of the current study. The proposed GP-DNN methodology may be extended to account for 3D effects, especially if well-screen depths may be combined with water level data to map the well observations as a function of latitude, longitude and depth. Notwithstanding the above limitation, one of the primary advantages of the GP-DNN model is the fast and analytical derivation of a statistically consistent posterior uncertainty model on long-term and seasonal groundwater trends informed by the lithological texture of the underlying aquifer layers. Informative probability distributions on key decision variables are a staple component for aiding decision making under uncertainty (\cite{caers11}, \cite{eidsvik15}). While the application was demonstrated specifically for the CV, similar data and uncertainty challenges exist in other hydrological basins where it should be possible to leverage latent space GPs for aiding uncertainty quantification.

It was shown earlier that the GP-DNN model did not significantly reduce the prior predictive uncertainty in the southern SJV. We identified two contributing factors that led to uninformative uncertainty quantification in this region: (1) highly variable hydrogeological heterogeneity, and (2) sparse well data. In the absence of sufficient information, the GP-DNN model correctly indicated that predictions were not reliable by putting large error bars on the posterior predictive mean in the southern SJV (\ref{fig:1dprofiles_GPDNN}). A specific novelty of the GP-DNN model is that the uncertainty predictions are driven by the hydrogeological heterogeneity, in addition to spatial proximity of data observations, as compared to the traditional kriging-based approaches which only account for the latter. This resulted in predictions of tight uncertainty intervals even with sparse well data in certain regions, for instance compare data density in central to southern region of Solano subbasin (Figure \ref{fig:cv_map}) with the predicted uncertainty intervals (Figure \ref{fig:1dprofiles_GPDNN}).

A limitation with GP-DNN posterior predictive samples (Figure \ref{fig:prior_realizations_S1_S2_intercept}) is the apparent loss of spatial continuity in the southern SJV, with large swings of the variables observed between spatially close locations. In addition to gathering more data which will constrain the posterior predictive uncertainty as discussed above, we propose two future improvement directions specific to the methodology. In the proposed GP-DNN formulation, Gaussian smoothing was performed in the latent space and not directly on the spatial coordinates. Potential loss of spatial continuity may be prevented by enforcing spatial regularization directly. A simple trick to achieve this will be to augment the latent space with spatial information. Specifically, the latent space may be specified as $\Tilde{\bm{x}}=\phi(\check{\bm{x}}$$;\bm{\theta})$, with the GP regression subsequently conducted in the augmented latent space ${[\bm{x}, \Tilde{\bm{x}}]}^T$. In this case, the kernel length scales along the spatial and latent coordinates will need to be carefully tuned. The GP kernel will be expected to suppress large deviations of water levels within the specified spatial length-scales.

While the GP-DNN model allows accounting for observational noise, it was assumed that the noise level is spatially homogeneous across the valley. This assumption may be violated in southern SJV given that the density of the well data samples is heavily skewed towards the northern part of the valley. This could potentially lead to higher levels of noise during fitting of the trend parameters in equation \ref{eq:linearReg1}. This limitation may be addressed by specifying the noise level to vary spatially with each location. Since the true noise level is unknown, the noise level will have to be considered as a parameter of the model. Within the formulation presented in section \ref{methodology}, it should be possible to derive the gradients of the loss function with respect to the noise parameters and train them end-to-end along with the neural network parameters. Spatially varying noise levels will provide the DNN flexibility to locally adjust the distances between the training locations in the latent space and prevent overfitting to any short correlations exhibited in the southern SJV data. 

As discussed in section \ref{latent space}, a common challenge associated with deep learning models is the difficulty associated in physically understanding how the input features affect the model outputs. In this paper, we approached model explainability by dimension reduction of the DNN estimated latent space and visualization of lithogical features along the latent reduced dimensions. This was effective in understanding how the DNN was effective in clustering together spatial locations with similar lithological characteristics. The reader is referred to the review paper by \cite{samek21} for an overview of other state-of-the-art methods available for explaining deep learning models such as interpretable local surrogate models, feature perturbation methods, and layer-wise relevance propagation. A general limitation of covariance based approaches is that they scale as $\mathcal{O}(n^3)$ with the number of training samples $n$. While this was not a computational bottleneck for our training dataset of 1550 wells, this might introduce significant computational burden for larger data sets. Addressing computational challenges of covariance matrix based algorithms is a well studied research area. \cite{rasmussen06}[Chapter~8] review several approximations methods that may be employed for applying GP-DNN to larger datasets.

\section{Conclusions}
A spatially continuous map of the groundwater level in CV is difficult to obtain due to the poor-quality of well data. In this paper, we proposed regression of sparse and noisy well data on features from a 3D lithological texture model of the CV aquifer system. We formulated a novel multivariate regression methodology that hierarchically leverages deep neural networks to morph the texture feature space into a latent space and Gaussian processes for non-parametric regression in the latent space. The proposed GP-DNN model provides a robust extension to traditional cokriging approach for modeling non-stationary data and augmenting uncertainty quantification with information from lithological features. We found that the GP-DNN model successfully captures non-stationary effects in the data by distending and squashing input distances in the latent space. The DNN was shown to be able to extract hydrogeologically explainable features from the data and the predictive uncertainty model was cross-validated to be statistically consistent with the empirical data distribution of 90 blind wells. Our results indicate that during 2015-2020 water levels in CV did not show appreciable recovery from the 2012-2015 drought in California. While the 2017 and 2019 wet years resulted in small and localized recovery of water levels, groundwater levels in August 2020 stayed mostly low in many areas of the valley. These results demonstrate promising applications of latent-space GP models to overcome data limitation challenges within hydrology and also have implications for refining our understanding of hydrologic connectivity in the context of groundwater recharge and drought recovery.

\section*{Open Research}
The CV lithologic texture data is available via the United States Geological Survey data release at \url{https://doi.org/10.5066/P9IZRO3V} (\cite{cvhm22}), and the CV digitial elevation model is available at \url{https://doi.org/10.5067/MEaSUREs/NASADEM/NASADEM_HGT.001} (\cite{nasadem21}). The CV well water level data is attributed to \cite{kim21}. The well data, processed as described in section \ref{data} along with the GP-DNN modeled water level trends and time series outputs, may be accessed through the Harvard Dataverse repository at \url{https://doi.org/10.7910/DVN/23TNJO} with Creative Commons Attribution 4.0 International license (\cite{pradhan24a}). Version v0.1.0 of Python software for GP-DNN regression, written using open source Tensorflow (\cite{tensorflow23}), NumPy (\cite{numpy20}) and Scipy (\cite{scipy20}) libraries, is preserved at \url{https://doi.org/10.5281/zenodo.13855361}, available via Creative Commons Attribution 4.0 International license (\cite{pradhan24b}). Normal score transformation method was performed using mGstat geostatistical MATLAB toolbox (\cite{hansen22}). All data analyses were performed using open source NumPy and Scipy Python libraries, while data visulaizations were conducted using open source Matplotlib Python library (\cite{matplotlib07}, \cite{matplotlib21}) and its Basemap extension. 

\acknowledgments
We thank Caltech's Resnick Sustainability Institute for funding the work presented and Professor Mark Simons and Dr. Neil Fromer for helpful discussions on groundwater modeling. Professor Venkat Chandrasekaran was supported in part by AFOSR grants FA9550-23-1-0204 and FA9550-22-1-0225, and by NSF grant DMS 2113724. Professor Andrew M. Stuart gratefully acknowledges support by the Air Force Office of Scientific Research under MURI award number FA9550-20-1-0358 (Machine Learning and Physics-Based Modeling and Simulation). We offer thanks to Dr. Kyongsik Yun and Dillon Holder for help with the processing of the data shown. We are also grateful towards Professor Tapan Mukerji whose expertise with geostatistics helped provide valuable insights supporting the presented research.
\bibliography{refs}

\appendix
\section{Construction of covariance matrices} \label{Appendix: Covariance}
The general anisotropic version of the stationary covariance kernel (equation \ref{eq:stationaryKernel}) is given as
$$
k(\bm{x},\bm{x}')= \sum_i K_{amp}^i k_{valid}^i\bigg(\sqrt{(\bm{x}-\bm{x}')^TL^{-1}(\bm{x}-\bm{x}')}\bigg)
$$
where $\bm{x} \in \mathbb{R}^n$ and $L=diag([l^2_1,l^2_2, \ldots, l^2_n]^T)$ is a diagonal matrix of the anisotropic length-scales along the input coordinates. The kernel length-scale may be related to the more traditional semivariogram range $r$, which is defined as the distance at which the semivariogram value reaches 95\% of the semivariogram sill value (\cite{goovaerts97}) as $l = \frac{r}{3}$. Note that the length-scales need to be specified for the baseline GP regression case. For simplicity, we assumed unit length-scales in the GP-DNN regression as the DNN can be expected to implicitly learn the scaling as part of the transformation $\phi(.)$.

The Matérn kernel used in equation \ref{eq:covKernel} is given as
$$
k_{Mat\Acute{e}rn}^{\nu}(d) = \frac{2^{1-\nu}}{\Gamma(\nu)}(  \sqrt{2\nu}d )^\nu K_\nu (\sqrt{2\nu}d), 
$$
where $d$ is the scaled distance between two locations under evaluation, $K_\nu$ is a modified Bessel function, and $\nu$ is a roughness parameter. As a result of the linear model of coregionalization, the amplitude matrix $K_{amp}$ is required to be positive semi-definite. $K_{amp}$ controls the variance of the components of the multivariate random process and their correlation coefficients. We assumed each individual process to have unit variance and normalized the well training data appropriately. The off-diagonal elements of $K_{amp}$ control the correlation coefficient between the processes since
$$
\rho_{a_i(\bm{x}), a_j(\bm{x})} = \frac{cov[a_i(\bm{x}), a_j(\bm{x})]}{\sigma_{a_i(\bm{x})}\sigma_{a_j(\bm{x})}} = K_{amp}^{i,j}k_{Mat\Acute{e}rn}^{\nu=2.5}(\bm{x},\bm{x}) = K_{amp}^{i,j},
$$
where $\rho$ denotes the correlation coefficient, $cov[.]$ is the covariance operator and $\sigma$ denotes standard deviation. 

Given training data samples $\{(\bm{x}_{\tau_i}, \bm{y}_{\tau_1}); i=1,\ldots,m\}$ such that $\bm{x}_{\tau} \in \mathbb{R}^n$ and $\bm{y}_{\tau} \in \mathbb{R}^d$, the $4m \times 4m$ covariance matrix
\begin{linenomath*}
\begin{equation}
    \label{eq:maternKernel}
K_{\tau\tau}=
\begin{bmatrix}
    K^{1,1}_{\tau\tau} & \ldots & K^{1,4}_{\tau\tau}\\
    \vdots & \ddots & \vdots \\
    K^{4,1}_{\tau\tau} & \dots & K^{4,4}_{\tau\tau}
\end{bmatrix},
\end{equation}
where,  $$ 
K^{i,j}_{\tau\tau}=
    \begin{bmatrix}
    k^{i,j}(\bm{x}_{\tau_1},\bm{x}_{\tau_1}) & \hdots 
    & k^{i,j}(\bm{x}_{\tau_1},\bm{x}_{\tau_m})\\
    \vdots & \ddots & \vdots\\
    k^{i,j}(\bm{x}_{\tau_m},\bm{x}_{\tau_1}) & \hdots & k^{i,j}(\bm{x}_{\tau_m},\bm{x}_{\tau_m})
    \end{bmatrix},\quad \forall i,j=1,\ldots,d.
    $$ 
    \end{linenomath*}
In the above,
$$
k^{i,j}(\bm{x}_{\tau_k},\bm{x}_{\tau_l}) = K_{amp}^{i,j}k_{Mat\Acute{e}rn}^{\nu=2.5}(\bm{x}_{\tau_k},\bm{x}_{\tau_l}).
$$

\section{Time series linear regression}
\label{Appendix: Time series regression}
The mathematical model capturing long-term and seasonal trends in well water level time series data is given in equation \eqref{eq:linearReg1}. We derive how the parameters $c$ may be estimated independently at each training location by linear regression. Equation \ref{eq:linearReg1} can be re-written as
    \begin{linenomath*}
    \begin{equation}
    \label{eq:linearReg2}
        \begin{gathered}
        u(\bm{x},t)=a_1(\bm{x})+a_2(\bm{x})t+
        a_3'(\bm{x})sin\bigg( \frac{2\pi t}{\lambda}\bigg) +
        a_4'(\bm{x})cos\bigg( \frac{2\pi t}{\lambda}\bigg),
         \end{gathered}
    \end{equation}
    \end{linenomath*}
where, $a_3'(\bm{x})=a_3(\bm{x})cos\big(a_4(\bm{x})\big)$ and $a_4'(\bm{x})=a_3(\bm{x})sin\big(a_4(\bm{x})\big)$. Given data $\{(t_j, u_{\bm{x}_{\tau_i},j}); j=1,\ldots,n_i\}$ at the $i^{th}$ well, equation \eqref{eq:linearReg2} may be used to specify the following system of linear equations
\begin{linenomath*}
    \begin{equation}
        \label{eq:linearReg3}
            \bm{u}_{\tau_i}=X_{\tau_i}\bm{a}_{\tau_i},
    \end{equation}
\end{linenomath*}
where, $\bm{u}_{\tau_i}$ is the $n_i \times 1$ vector of water levels, $X_{\tau_i}$ is the $n_i \times 4$ feature matrix and $\bm{a}_{\tau_i}=\big[a_1(\bm{x}_{\tau_i}), a_2(\bm{x}_{\tau_i}), a_3'(\bm{x}_{\tau_i}), a_4'(\bm{x}_{\tau_i}) \big]^T$ is the trend parameter vector to be estimated. The least-squares solution to equation \eqref{eq:linearReg3} is given as
\begin{linenomath*}
\begin{equation}
    \label{eq:linInv1}
     \bm{a}_{\tau_i}=\big(X_{\tau_i}^T X_{\tau_i}\big)^{-1}X_{\tau_i}^T \bm{u}_{\tau_i}.
\end{equation}
\end{linenomath*}

We solve equation \eqref{eq:linInv1} independently at each well location to obtain estimates of the corresponding parameter vector $\bm{a}_{\tau_i}$. The seasonal amplitude $a_3$ and phase delay $a_4$, computed as
$$
    a_3=\sqrt{a_3'^2+a_4'^2}\ \textrm{and}\ a_4=\arctantwo(a_3',a_4')
$$
respectively. 

\section{Training and tuning of regression models}
\label{Appendix: trainingTuning}
\subsubsection{Baseline GP regression} \label{Appendix: trainingGPspatial}
In this case, the regression is performed in the geospatial coordinate space $\bm{x}$. The posterior predictive distribution may be analytically derived as shown in equation \ref{eq:cokriging}. The list of hyper-parameters is shown in Table \ref{table:hyperParamGPspatial}. For each hyper-parameter, we specify a range of possible values it may assume. In general, available data at training wells were used as a guide to specify the support of the hyper-parameters. For instance, to estimate the ranges for length-scales $l_1$ and $l_2$ along latitude and longitude coordinates, empirical variogram analysis (\cite{goovaerts97}) with well data indicated that length-scales ranged roughly between 5-15 miles. To account for the uncertainty in the empirical estimates, we considered lower and upper bounds of 3 and 75 miles respectively. The other hyper-parameters considered are related  to the covariance kernel specification, i.e., $K_{amp}$ and $\Sigma_n$. The diagonal elements of the Matérn kernel amplitude matrix $K_{amp}$, defining the signal variances of $a_i, i=1,\ldots,4$, are taken to be 1 given that we normalized all the training, validation and test data to have unit variance. Uncertainty in the diagonal entries of the covariance matrix is modeled through the $\Sigma_n$ described below. The off-diagonal entries of the  $K_{amp}^{i,j}, i \neq j,$ capture the correlation coefficients between the trend parameters (\ref{Appendix: Covariance}). Note that there are only six free off-diagonal elements, since $K_{amp}$ is required to be positive semi-definite by definition. We consider higher noise variances for $a_3$ and $a_4$ since estimating the sinusoidal phase parameter $a_4'$ from the sparse well time series will typically be more difficult than intercept, slope and sinusoidal amplitude, resulting in noisier estimates of $a_3$ and $a_4$ from equation \ref{eq:linInv1}). Hyper-parameter values tuned by cross-validation are shown in Table \ref{table:hyperParamGPspatial}.  

\begin{table}[!ht]
\begin{center}
\begin{tabular}{ |c|c|c| }
\hline
    \textbf{Hyper-parameter} & \textbf{Parameter range} & \textbf{Tuned value} \\ 
    \hline
    Length-scale $l_1$ & [3 miles, 75 miles] & 12.13 miles \\ 
    \hline
    Length-scale $l_2$ & [3 miles, 75 miles] & 7.11 miles\\ 
    \hline
    Noise variance $\sigma_{n_1}^2$ & [0.2, 0.5] & 0.23 \\ 
    \hline
    Noise variance of $\sigma_{n_2}^2$ & [0.2, 0.7] & 0.61 \\ 
    \hline
    Noise variance of $\sigma_{n_3}^2$ & [0.4, 0.98] & 0.43 \\ 
    \hline
    Noise variance of $\sigma_{n_4}^2$ & [0.4, 0.98] & 0.70 \\ 
    \hline
    Kernel amplitude $K_{amp}^{1,2}$ & [-0.7, -0.25] & -0.41\\
    \hline
    Kernel amplitude $K_{amp}^{1,3}$ & [0.01, 0.2]  & 0.16\\ 
    \hline
    Kernel amplitude $K_{amp}^{1,4}$ & [-0.4, -0.05] & -0.37\\ 
    \hline
    Kernel amplitude $K_{amp}^{2,3}$ & [-0.15, 0.15] & 0.10\\ 
    \hline
    Kernel amplitude $K_{amp}^{2,4}$ & [-0.05, 0.15] & -0.02\\ 
    \hline
    Kernel amplitude $K_{amp}^{3,4}$ & [-0.9, -0.4] & -0.80\\ 
\hline
\end{tabular}
\end{center}
\caption{Hyper-parameter tuning details for baseline GP regression.}
\label{table:hyperParamGPspatial}
\end{table}

\subsubsection{Hierarchical GP-DNN regression} \label{Appendix: GP-DNN training}
In this case, the prior predictive distribution is specified using the hierarchical model posited in equations \ref{eq:hierarchicalGP1} and \ref{eq:hierarchicalGP2}. The DNN in the bottom layer may be parameterized through several hidden layers, each of which constitutes of a number of neurons with trainable weight and bias parameters $\bm{\theta}$ to yield multivariate outputs. We treat the number of hidden layers, neurons and outputs as hyper-parameters taking values in the ranges shown in Table \ref{table:hyperParamGPextended}. At the top-level is a GP model, regressed in the space of DNN outputs $\Tilde{\bm{x}}$. The posterior predictive distribution on prediction variables may then be derived as shown in equation \ref{eq:likelihood_metric}. Given above model specification, parameters $\bm{\theta}$ will be optimized by minimizing the negative log-likelihood of the GP marginal distribution, and hyper-parameters tuned by cross validation with the negative log-likelihood of the validation data under the GP posterior predictive distribution. To limit overfitting, we considered dropout (\cite{srivastava2014}) and $\ell_2$-regularization, with dropout regularization found to be largely ineffective (Table \ref{table:hyperParamGPextended}). Based on cross-validation, the optimal DNN architecture was found to consist of 2 hidden layers with 33 neurons in each layer. The dimension of the latent space was tuned to be 12. In addition to the DNN hyper-parameters, we also consider hyper-parameters related to the GP Mátern kernel and data noise. For simplicity, we assumed unit length-scales $\bm{l} \in \mathbb{R}^{p\times1}$. 

\begin{figure}[!ht]
    \centering
    \includegraphics[scale=0.25]{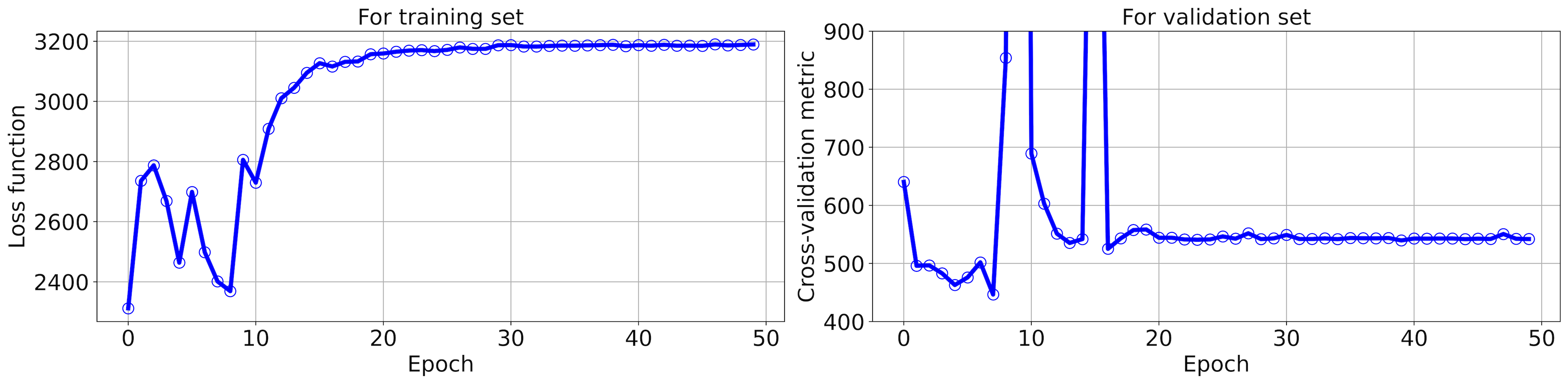}
    \caption{GP-DNN training history across 100 epochs. (Left) Negative of marginal log-likelihood loss function (R.H.S. of equation \ref{eq:logLikelihood} without the constant term) evaluated on the training set is plotted along the y-axis. (Right) Negative of predictive posterior log-likelihood evaluated on the validation set is plotted along the y-axis} 
    \label{fig:train_history}
\end{figure}

The training procedure is shown in Algorithm \ref{alg:GPextendedTrain}. The training algorithm takes as inputs the training data feature matrix $\Hat{X}_{\tau}\in \mathbb{R}^{m \times (n+2)}$, where $m=1550$ is the number of training examples and $n=39$ is the dimension of the hydrogeological feature space, and the target vector $\bm{y}_\tau \in \mathbb{R}^{4m\times1}$. In Figure \ref{fig:train_history}, we show the variability of the training set loss function across 100 training epochs. Note that the loss function in equation \ref{eq:logLikelihood} comprises of the data fit and model complexity terms (see similar discussion for the posterior predictive distribution in section \ref{likelihood_interpretation}). As the training progresses, the DNN model parameters $\bm{\theta}$ are expected to become increasingly complex to overfit to the training data leading to larger values of the model complexity term. This is clearly observed in the increasing trend of the loss function beyond the tenth epoch, driven by increasing model complexity term. To ensure that the DNN model parameters may generalize to datasets other than the training set, we enforce early stopping of the training based on the cross-validation metric of negative log-likelihood of the predictive posterior distribution evaluated on the validation set. In the right subplot of Figure \ref{fig:train_history}, we show the behavior of validation set cross-validation metric across 100 training epochs. The best validation metric is attained at the seventh epoch, hence the DNN parameters $\bm{\theta}$ from this epoch are employed for all subsequent model predictions.
\begin{table}[!ht]
\begin{center}
\begin{tabular}{ |c|c|c| }
\hline
\textbf{Hyper-parameter} & \textbf{Parameter range} & \textbf{Tuned value} \\
    \hline
    Number of hidden layers & \{1, 2, 3\} & 2 \\
    \hline
    Number of neurons in hidden layers & \{30, 31, \ldots, 130\} & 33 \\
    \hline
    Number of DNN output nodes & \{1, 2, \ldots, 30\} & 12 \\
    \hline
    Use dropout & \{True, False\} & False \\
    \hline
    \makecell{Learning rate for DNN training \\ with Adam optimization algorithm} & [0.001, 0.5] & 0.28 \\
    \hline
    Weight of $\ell_2$-regularization & [0.001, 10] & 2.91 \\
    \hline
    Noise variance $\sigma_{n_1}^2$ & [0.2, 0.5] & 0.28 \\ 
    \hline
    Noise variance of $\sigma_{n_2}^2$ & [0.2, 0.7] & 0.54 \\ 
    \hline
    Noise variance of $\sigma_{n_3}^2$ & [0.4, 0.98] & 0.93 \\ 
    \hline
    Noise variance of $\sigma_{n_4}^2$ & [0.4, 0.98] & 0.85 \\ 
    \hline
    Kernel amplitude $K_{amp}^{1,2}$ & [-0.7, -0.25] & -0.42\\
    \hline
    Kernel amplitude $K_{amp}^{1,3}$ & [0.01,0.2]  & 0.11\\ 
    \hline
    Kernel amplitude $K_{amp}^{1,4}$ & [-0.4,-0.05] & -0.28\\ 
    \hline
    Kernel amplitude $K_{amp}^{2,3}$ & [-0.05,0.15] & -0.06\\ 
    \hline
    Kernel amplitude $K_{amp}^{2,4}$ & [-0.05,0.15] & 0.03 \\ 
    \hline
    Kernel amplitude $K_{amp}^{3,4}$ & [-0.9,-0.4] & -0.47\\ 
\hline
\end{tabular}
\end{center}
\caption{Hyper-parameter tuning details for GP-DNN regression.}
\label{table:hyperParamGPextended}
\end{table}

\RestyleAlgo{ruled}
\begin{algorithm}[!ht]
\caption{Algorithm for training of the GP-DNN hierarchical model}\label{alg:GPextendedTrain}
\KwIn{
\par
Training data $\{\Hat{X}_\tau, \bm{y}_\tau\}$ and validation data $\{\Hat{X}_*, \bm{y}_*\}$\\ 
Hyper-parameters listed in Table \ref{table:hyperParamGPextended}, length-scales $\bm{l}=\textbf{1}$\\ 
Parameters $\bm{\theta}$ initialized using the Glorot uniform distribution (\cite{glorot2010})\\ 
$n_{epoch}$: number of training epochs.}
\vspace{0.25cm}
\For{$i=1,\ldots,n_{epoch}$}{
        \nl Compute $\Tilde{\bm{x}}_{\tau_i} = \phi(\hat{\bm{x}}_{\tau_i};\bm{\theta}) \, \forall \, i \in \{1,\ldots,m\}$, and $\Tilde{X}_\tau$.\\
        \nl Formulate $\Tilde{K}_{\tau\tau}$ as given by equation \ref{eq:maternKernel}.\\
        \nl Compute the negative log-likelihood of the GP marginal distribution $\textrm{log} \ p(\bm{y}_\tau|\hat{X}_\tau;\bm{\theta})$ as given in equation \ref{eq:logLikelihood} and save to training history.\\
        \nl Derive $\frac{\partial}{\partial \theta_k} p(\bm{y}_\tau|\hat{X}_\tau;\bm{\theta}), \forall \, k$ using equation \ref{eq:Derivative}, where $\frac{\partial \Tilde{K}_{\tau\tau}}{\partial \theta_k}$ is calculated by the back-propagation algorithm.\\
        \nl Use the Adam stochastic optimization algorithm (\cite{kingma2015}) to update $\bm{\theta}$ and save to training history.\\
        \nl Estimate the mean and covariance matrix of the posterior predictive distribution $\bm{a_*}|\bm{y}_\tau,\hat{X}_\tau,\hat{X}_*$ (similar to equation \ref{eq:cokriging}).\\
        \nl Evaluate the cross-validation statistic according to equation \ref{eq:likelihood_metric} and save to training history.
}
\vspace{0.25cm}
\KwOut{
\par
Trained $\bm{\theta}$: set to that estimated after epoch with highest cross-validation metric\\
Training history: evolution of likelihood of training data, cross-validation metric and $\bm{\theta}$ across all epochs.}
\end{algorithm}

\end{document}